\documentclass[notitlepage,a4paper,aps,prd,onecolumn,superscriptaddress,nofootinbib,groupedaddress]{revtex4-2}

\usepackage{amsmath}
\usepackage{amsfonts,color}
\usepackage{amsthm}
\usepackage{ulem}
\usepackage{empheq}
\usepackage{amssymb,float}
\usepackage{booktabs,multirow}
\usepackage{titlesec}
\usepackage{hyphenat}
\usepackage{ragged2e}
\usepackage{amsmath}
\usepackage{accents}
\usepackage[utf8]{inputenc}
\usepackage{color}
\usepackage{hyperref}
\usepackage{enumitem}
\usepackage{tikz}
\usetikzlibrary{shapes.geometric}
\usetikzlibrary{arrows.meta,arrows}

\titleformat{\paragraph}
{\normalfont\normalsize\bfseries}{\theparagraph}{1em}{}
\titlespacing*{\paragraph}
{0pt}{3.25ex plus 1ex minus .2ex}{1.5ex plus .2ex}
\newcommand{\lc}[1]{\accentset{\circ}{#1}}
\allowdisplaybreaks
\hypersetup{
    colorlinks=true,
    linkcolor=blue,
    filecolor=magenta,      
    citecolor=blue
}
\begin{document}

\title{
Degrees of Freedom of New General Relativity I:\\
Type 2, Type 3, Type 5, and Type 8
%:\\
%Physically interesting Types
}
\author{Kyosuke TOMONARI}
\email{ktomonari.phys@gmail.com}
\affiliation{Department of Physics, Institute of Science Tokyo, 2-12-1 Ookayama, Meguro-ku, Tokyo 152-8551, Japan\\}
\affiliation{Interfaculty Initiative in Information Studies, Graduate School of Interdisciplinary Information Studies, The University of Tokyo, 7-3-1 Hongo, Bunkyo-ku, Tokyo 113-0033, Japan}
\author{Daniel Blixt}
\email{daleblixt@gmail.com}
\affiliation{Scuola Superiore Merdionale, Largo S. Marcellino 10, I-80138, Napoli, Italy}

\begin{abstract}
We investigate the degrees of freedom of new general relativity. This theory is a three-parameter theory and is classified into nine irreducible types according to the rotation symmetry of $SO(3)$ on each leaf of ADM-foliation. In this work, we focus on unveiling the degrees of freedom of the physically interesting types of NGR: Type 2, Type 3, Type 5, and Type 8, which contain the gravitational propagating degrees of freedom. First, we revisit the theory based on the gauge approach to gravity and reformulate the Lagrangian of the theory. Second, we review the irreducible decomposition of the theory while focusing on the Hamiltonian and the primary constraints in each type. Third, we perform the Dirac-Bergmann analysis to unveil the degrees of freedom of the theory in the case of Type 2, Type 3, Type 5, and Type 8. We find a novel new behavior of constraints in Type 8, which is classified as second-class but not to determine any Lagrange multipliers and to provide the gauge invariance of the theory under the satisfaction of a specific condition of the multipliers. The degrees of freedom of Type 2, Type 3, and Type 5 are unveiled as six, five, and seven, respectively. The degrees of freedom of Type 8 is either four under a specific condition to the Lagrange multipliers or six in the generic case. Finally, we conclude this work with five future perspectives.
\end{abstract}

\maketitle

\section{\label{01}Introduction}
General Relativity (GR) is the most successful classical theory of gravity to describe the wide range of gravitational phenomena by applying pseudo-Riemannian geometry based on the local Lorentz invariance, the diffeomorphism symmetry, and Einstein's equivalence principle. In physical perspectives, however, there is no reason to restrict our theories to this particular geometry. Historically, Einstein reconstructed GR in an alternative manner using different geometry based purely on torsion instead of curvature, which is known as teleparallel gravity~\cite{Einstein1928}. For a detailed review of teleparallel gravity, see Ref.~\cite{Bahamonde:2021gfp} and the Refs. therein. In modern perspectives, it is well-known that GR has its equivalent formulation in which gravitation is treated with the torsion (Teleparallel Equivalent to GR: TEGR) and/or the non-metricity (General/Symmetric Teleparallel Equivalent to GR: GTEGR or STEGR, respectively) instead of the curvature up to boundary terms, labeled as geometrical Trinity of Gravity (ToG)~\cite{Nester:1998mp,BeltranJimenez:2019esp,Heisenberg:2018vsk,BeltranJimenez:2019odq}. These two theories assume that the general curvature is vanishing. In more generic perspectives, ToG is a set of specific classes in the so-called Metric-Affine gauge theories of Gravity (MAG), disciplined by the property of gauge symmetries~\cite{Utiyama:1956sy,Kibble:1961ba,Ivanenko:1983fts}. For a detailed review on MAG, see Refs.~\cite{Hehl:1994ue} and the Refs. therein. 

GR and the standard model of elementary particles are fundamental theories for establishing modern cosmology~\cite{Dodelson:2003ft,Mukhanov:2005sc,Weinberg:2008zzc}. Observations have unveiled the new perspectives in modern cosmology, such as the necessity of inflation~\cite{Planck:2018vyg,Tsujikawa:2003jp,Vazquez:2018qdg}, the existence of dark matter~\cite{Freese:2008cz,Billard:2021uyg,Planck:2018vyg}, the late-time acceleration of the universe (or the necessity of dark energy)~\cite{Planck:2018vyg,SupernovaSearchTeam:1998fmf,SupernovaCosmologyProject:1998vns}, and most recently the tension in the Hubble constant~\cite{Planck:2018vyg,H0LiCOW:2019pvv,Riess:2019cxk,Schoneberg:2022ggi,ACT:2023kun}. These issues suggest that the fundamental theories of modern cosmology would suffer from some difficulties. In fact, first, GR cannot explain the gravitational phenomena in quantum perspectives due to the fault of renormalizability~\cite{Wald:1984rg,Hawking:1973uf,Weinberg:1972kfs}. This gives rise to the difficulty in describing the origin of inflation. Second, GR provides a consistent explanation of the late-time acceleration of the universe by {\it a posteriori} inducing the so-called cosmological constant into Einstein's field equation, but it has not been clarified the origin of the term~\cite{Peebles:2002gy,Padmanabhan:2002ji,Carroll:2000fy}. Namely, this is just a phenomenological explanation, and the term is not derived from the first principle. Third, GR has no mechanism to reconcile the Hubble-tension due to that the theory treats gravitation in a constant strength, {\it i.e.,} the Newtonian constant~\cite{Wald:1984rg,Hawking:1973uf,Padmanabhan:2010zzb}. One of the approaches to challenge such issues is to reconsider the fundamental theory of gravity, {\it i.e.,} GR, on the ground of MAG frameworks. MAG and/or its extension/modification can give a perspective to explain these issues \cite{Dwivedi:2024okk}. In particular, the non-linear extension of MAG theories in the same manner as $f(R)$-gravity is remarkable for approaching these issues. For a detailed review on the extended theories, see Refs.~\cite{Nojiri:2006ri,DeFelice:2010aj,Cai:2015emx,Heisenberg:2023lru,Zhao:2021zab} and Refs. therein. The extended theories provide the well-behaved inflation models~\cite{Buchdahl:1970ldb,Rezazadeh:2017edd,Capozziello:2022tvv,Nojiri:2024zab,Nojiri:2024hau,Gamonal:2020itt,Capozziello:2024lsz} in the high precision to the recent observations given by Planck 18~\cite{Planck:2018nkj}. Furthermore, the extended theories {\it a priori} contain an effective cosmological constant and gravitational constant in their field equations, and these effective constants can explain the late-time acceleration of the universe~\cite{Tsujikawa:2010sc,Bamba:2010wb,Zubair:2015opa,Bahamonde:2017ize,Solanki:2022ccf,Nojiri:2024hau,Nojiri:2024zab} and reconcile the Hubble-tension~\cite{DiValentino:2021izs,Heisenberg:2022gqk}, respectively. These novel features are ascribed to the inherent extra Degrees of Freedom (DoF) in the theories. To identify the novel DoF, the Dirac-Bergmann analysis is valid~\cite{Dirac:1950pj,Dirac:1958sq,Bergmann:1949zz,BergmannBrunings1949,Bergmann1950,Anderson:1951ta}.

The investigation of the structure formation of the universe is also one of the significant issues in modern cosmology~\cite{Dodelson:2003ft,Mukhanov:2005sc,Weinberg:2008zzc,Matsubara:2022ohx,Matsubara:2022eui,Matsubara:2023avg,Matsubara:2024sqn}. To approach this issue, we employ the linear perturbation theory around the flat and non-flat Friedmann–Lemaître–Robertson–Walker (FLRW) spacetime~\cite{Bernardeau:2001qr,Malik:2008im}. However, in the extension/modification of the theories of gravity, a discrepancy in the number of the linear DoF in the perturbation and that of the non-linear DoF of a given theory generically occurs. The discrepancy is called the `{\it strong coupling}' around the background spacetime chosen for the perturbation in advance. For a detailed description, see Sec. IV in Ref.~\cite{Bahamonde:2024zkb}. Perturbation theories that suffer from this issue would not healthily predict physical phenomena due to the lack of the DoF existing in the origin theory. The extension/modification of MAG also encounters this issue. For example, see Refs.~\cite{Bahamonde:2020lsm,Bahamonde:2022ohm,Aoki:2023sum,Tomonari:2023wcs,Bahamonde:2024zkb}. To investigate whether the issue exists in a given theory or not, we have to unveil the non-linear DoF of the theory. Again, the Dirac-Bergmann analysis plays a crucial role in investigating this subject. 

In recent years, New General Relativity (NGR)~\cite{Hayashi:1979qx} and Newer General Relativity (Newer GR)~\cite{BeltranJimenez:2017tkd} have begun to gather attention as candidates for the new extension of MAG to reconcile the difficulties in modern cosmology. NGR and Newer GR are the three- and five-parameter family extensions of TEGR and CGR (STEGR in the coincident gauge~\cite{BeltranJimenez:2017tkd,BeltranJimenez:2022azb}), respectively. These theories are composed of nine independent types based on the irreducibility of canonical momenta in $SO(3)$-rotation in four-dimensional spacetime~\cite{Blixt:2018znp}. Most recently, the linear perturbation around the Minkowski background spacetime of NGR is performed, and the linear DoF and ghost-free conditions in each type of NGR are clarified~\cite{Bahamonde:2024zkb}. However, the nonlinear DoF that is unveiled by performing the Dirac-Bergmann analysis has not been fully unveiled yet.\footnote{It is known that Type 6 (TEGR) has two DoF as expected~\cite{Maluf:2000ag,Blagojevic:2000qs,Ferraro:2016wht}, and Type 1 (generic theory) has eight DoF~\cite{Mitric:2019rop}, Type 2 was analyzed in~\cite{Cheng:1988zg}, having six DoF only spelled out in a master thesis mentioned in this article.} 
Clarifying the number is very important for the application of NGR to modern cosmology, as explained above. In this paper, we perform the Dirac-Bergmann analysis of NGR to investigate the existence of strong couplings against the background spacetime, focusing on the physically interesting types: Type 2, Type 3, Type 5, and Type 8 since these types contain the propagating gravitational modes~\cite{Bahamonde:2024zkb}. 

The construction of the current paper is given as follows. In Sec.~\ref{02}, we reconstruct NGR based on the gauge approach to gravity, in which the Weitzenb\"{o}ck gauge is imposed and the non-metricity tensor automatically vanishes on the ground of the gauge invariant characteristics. This formulation makes the analysis simple. Then, the formulation of NGR in the $SO(3)$-irreducible manner is reviewed. In Sec.~\ref{03}, the Dirac-Bergmann analysis on each type of the theory is performed. In the current paper, we focus on Type 2, Type 3, Type 5, and Type 8. We unveil that the DoF of Type 2, Type 3, Type 5, and Type 8 are six, five, seven, and either four under a specific condition of Lagrange multipliers or six in a generic case, respectively. In particular, in the analysis on Type 8, we find a novel new behavior of second-class constraint densities. Namely, a specific condition of Lagrange multipliers allows the theory to be gauge invariant with respect even to the second-class constraint densities. Finally, in Sec.~\ref{04}, we summarize this work and give future perspectives. 

Throughout this paper, we use units with $\kappa=c^{4}/16\pi G_{N}:=1$. In the Dirac-Bergmann analysis, we denote ``$\approx$'' as the weak equality~\cite{Dirac:1950pj,Dirac:1958sq}. For quantities computed from the Levi-Civita connection, we use an over circle on top, whereas, for a general connection, tildes are introduced. Also, Greek indices denote spacetime indices, whereas Capital Latin ones denote the internal-space indices. Small Latin letters denote the spatial indices in the ADM-foliation~\cite{Deser:1959vvc,Arnowitt:1960es,Arnowitt:1962hi}.

\section{\label{02}Revisiting TEGR and NGR}
\subsection{\label{02:01}Gauge approach to TEGR and NGR}
Teleparallel theories of gravity is a set of special classes in more generic theory, so-called Metric-Affine gauge theories of Gravity (MAG)~\cite{Bahamonde:2021gfp}, and MAG is formalized based on the framework of gauge approach to gravity~\cite{Hehl:1994ue}. TEGR is a special class of teleparallel theories of gravity and NGR is an extension of TEGR using three free parameters.

Gauge approach to gravity demands two vector bundles~\cite{Tomonari:2023wcs,Tomonari:2023ars}: a tangent bundle $(T\mathcal{M}\,,\mathcal{M}\,,\pi)$ and an internal bundle $(\mathcal{V}\,,\mathcal{M}\,,\rho)$, where $\mathcal{M}$ is a spacetime manifold with dimension $n+1$, $\pi$ is an onto map from $T\mathcal{M}$ to $\mathcal{M}$, and $\rho$ is an onto map from $\mathcal{V}$ to $\mathcal{M}$. The total space of the internal bundle, $\mathcal{V}$, is called merely an internal-space, and in usual formulation, it is taken to be $\mathcal{M}\times\mathbb{R}^{n+1}$. In the current paper, we obey this ordinary choice of internal bundle. Then we introduce a frame field $\mathbf{e}\,:\,\mathcal{M}\times\mathbb{R}^{n+1}\rightarrow T\mathcal{M}$. In the component form, for a basis $\zeta_{A}$ on $\left.\mathcal{M}\times\mathbb{R}^{n+1}\right|_{U}$, where $U$ is an open set of $\mathcal{M}$, we can express the frame field $\mathbf{e}$ as follows: $e_{A}(p)=\mathbf{e}(p)(\zeta_{A})=e_{A}{}^{\mu}(p)\partial_{\mu}$, where $p\in U$. Note here the local property: $\left.\mathcal{M}\times\mathbb{R}^{n+1}\right|_{U}\simeq \left.T\mathcal{M}\right|_{U}$, in particular $\left.\mathcal{M}\times\mathbb{R}^{n+1}\right|_{p\in U}=\mathbb{R}^{n+1}\simeq\left.T\mathcal{M}\right|_{p\in U}$. The co-frame field of $\mathbf{e}$ is then defined as the inverse map of $\mathbf{e}$ as follows: $\mathbf{e}^{-1}\,:\,\left.T\mathcal{M}\right|_{U}\rightarrow\left.\mathcal{M}\times\mathbb{R}^{n+1}\right|_{U}$. In the component form, for the dual basis of $\zeta_{A}$, {\it i.e.,} $\zeta^{A}$, we have $\theta^{A}(p)=(\mathbf{e}^{-1})^{*}(p)(\zeta^{A})=\theta^{A}{}_{\mu}(p)dx^{\mu}$, where $(\mathbf{e}^{-1})^{*}$ denotes the pullback of $\mathbf{e}^{-1}$ and $p\in U$. Remark here that the inverse map of the frame field can be defined only in a local region of the spacetime. The frame field and co-frame field components, $e_{A}{}^{\mu}$ and $\theta^{A}{}_{\mu}$, on a local region $U$ satisfy the following properties: $e_{A}{}^{\mu}\,\theta^{A}{}_{\nu}=\delta^{\mu}{}_{\nu}$ and $e_{A}{}^{\mu}\theta^{B}{}_{\mu}=\delta^{A}{}_{B}$. Using these ingredients, a metric $g=g_{\mu\nu}dx^{\mu}\otimes dx^{\nu}$ on $\mathcal{M}$ and a metric $g=\eta_{AB}\zeta^{A}\otimes\zeta^{B}$ on $\mathcal{M}\times\mathbb{R}^{n+1}$ are related by 
\begin{equation}
    g_{\mu\nu}=\theta^{A}{}_{\mu}\theta^{B}{}_{\nu}\eta_{AB}\,\quad \eta_{AB}=e_{A}{}^{\mu}e_{B}{}^{\nu}g_{\mu\nu}\,.
    \label{relation between spacetime and internal space metric}
\end{equation}
Note here that the metric, $\eta_{AB}$, should be determined as a gauge in the internal-space. In the current paper, we set $\eta_{AB}$ as the Minkowskian metric. 

In MAG, the affine connection $\tilde{\Gamma}^{\rho}{}_{\mu\nu}$ in the spacetime $\mathcal{M}$ and the connection 1-form $\omega^{A}{}_{B\mu}$ in the internal-space $\mathcal{M}\times\mathbb{R}^{n+1}$ are related by the Weitzenb\"{o}ck connection as follows~\cite{Weitzenboh1923}:
\begin{equation}
    \tilde{\Gamma}^{\rho}_{\mu\nu} = e_{A}{}^{\rho}\,\partial_{\mu}\theta^{A}{}_{\nu} + e_{A}{}^{\rho}\,\theta^{B}{}_{\mu}\,\omega^{A}{}_{B\nu}\,.
\label{Weitzenbock connection}
\end{equation}
Remark that this relation assumes that the local property $\left.\mathcal{M}\times\mathbb{R}^{n+1}\right|_{U}\simeq \left.T\mathcal{M}\right|_{U}$ holds. It allows us to compute the covariant derivative of the co-frame field component, $e^{i}{}_{\mu}$, as follows: $\mathcal{D}_{\nu}\theta^{A}{}_{\mu}=\partial_{\nu}\theta^{A}{}_{\mu}-\tilde{\Gamma}^{\rho}{}_{\nu\mu}\theta^{A}{}_{\rho}+\omega^{A}{}_{B\nu}\theta^{B}{}_{\mu}$. Applying the Weitzenb\"{o}ck connection, the frame field postulate automatically holds: $\mathcal{D}_{\nu}\theta^{A}{}_{\mu}=0$. In a generic affine connection, a Lie group action to the co-frame field provides the attribute of internal-space symmetry at each spacetime point to our theories of gravity in the usual sense of gauge theory. Namely, a co-frame field transformation $\theta^{A}{}_{\mu}\rightarrow \theta'^{A}{}_{\mu}=\Lambda^{A}{}_{B}\theta^{B}{}_{\mu}$, where $\Lambda^{A}{}_{B}\in G$ and $G$ is a Lie group, leads to
\begin{equation}
    \mathcal{D}_{\mu}\theta'^{A}{}_{\nu}=\Lambda^{A}{}_{B}\mathcal{D}_{\mu}\theta^{B}{}_{\nu}
\label{gauge transformation of internal space derivative}
\end{equation}
on the ground of the transformation of the connection 1-form component as follows: $\omega^{A}{}_{B\mu}\rightarrow \omega'^{A}{}_{B\mu}=(\Lambda^{-1})^{A}_{\ C}\partial_{\mu}\Lambda^{C}{}_{B}+(\Lambda^{-1})^{A}{}_{C}\Lambda^{D}{}_{B}\omega^{C}{}_{D\mu}$. In the current paper, we can set this Lie group $G$ as the Lorentz group $SO(1\,,n)$ in a consistent manner. Therefore, if the Weitzenb\"{o}ck connection, Eq.~(\ref{Weitzenbock connection}), holds in a specific frame $\theta^{A}{}_{\mu}$, then so does in another frame $\theta'^{A}{}_{\mu}=\Lambda^{A}{}_{B}\theta^{B}{}_{\mu}$. In particular, we can always take the so-called Weitzenb\"{o}ck gauge~\cite{Adak:2005cd,Adak:2006rx,Adak:2008gd,Adak:2011ltj}
\begin{equation}
    \omega^{A}{}_{B\mu}=0\,.
\label{Weitzenbock gauge}
\end{equation}
Therefore, in a generic frame we have
\begin{equation}
    \tilde{\Gamma}^{\rho}_{\mu\nu} = e_{A}{}^{\rho}\,\partial_{\mu}\theta^{A}{}_{\nu}\,,\quad \omega'^{A}{}_{B\mu} = (\Lambda^{-1})^{A}_{\ C}\partial_{\mu}\Lambda^{C}{}_{B}\,.
\label{Affine connection and connection 1-form in Weitzenbock gauge}
\end{equation}
The first formula above holds in any frame choice by virtue of Eq.~(\ref{gauge transformation of internal space derivative}). In this specific gauge choice, the teleparallel condition is automatically satisfied, {\it i.e.} $\tilde{R}^{\sigma}{}_{\mu\nu\rho}=2\partial_{[\nu}\tilde{\Gamma}^{\sigma}{}_{\rho]\mu}+2\tilde{\Gamma}^{\sigma}{}_{[\nu|\lambda|}\tilde{\Gamma}^{\lambda}{}_{\rho]\mu}=0$, which is independent of choosing frames. While we have the relation
\begin{equation}
    \tilde{R}^{\sigma}{}_{\mu\nu\rho}=\lc{R}^{\sigma}{}_{\mu\nu\rho}+2\lc{\nabla}_{[\nu}N^{\sigma}{}_{\rho]\mu}+2N^{\sigma}{}_{[\nu|\lambda|}N^{\lambda}{}_{\rho]\mu}
\label{Generic curvature tensor}
\end{equation}
for a distorsion tensor $N^{\rho}{}_{\mu\nu}=\tilde{\Gamma}^{\rho}{}_{\mu\nu}-\lc{\Gamma}^{\rho}{}_{\mu\nu}$, where $\lc{\Gamma}^{\rho}{}_{\mu\nu}$ is the Levi-Civita connection. In MAG, the distorsion tensor is decomposed into the contorsion, $K^{\rho}{}_{\mu\nu}$, and the disformation, $L^{\rho}{}_{\mu\nu}$, as follows:
\begin{equation}
    N^{\rho}{}_{\mu\nu}=K^{\rho}{}_{\mu\nu}+L^{\rho}{}_{\mu\nu}\,,
\label{distorsion}
\end{equation}
where 
\begin{equation}
    K^{\rho}{}_{\mu\nu}=\frac{1}{2}T^{\rho}{}_{\mu\nu}+T_{(\mu\,\,\,\nu)}^{\,\,\,\,\,\rho}\,,\quad L^{\rho}{}_{\mu\nu}=\frac{1}{2}Q^{\rho}{}_{\mu\nu}-Q_{(\mu\,\,\,\nu)}^{\,\,\,\,\,\rho}\,,
\label{contorsion and disformation}
\end{equation}
respectively and, the torsion $T^{\rho}{}_{\mu\nu}$ and the non-metricity $Q^{\rho}{}_{\mu\nu}$ are defined by
\begin{equation}
    T^{\rho}{}_{\mu\nu}=e_{A}{}^{\rho}T^{A}{}_{\mu\nu}=2e_{A}{}^{\rho}\partial_{[\mu}\theta^{A}{}_{\nu]}\,,\quad Q^{\rho}{}_{\mu\nu}=g^{\rho\lambda}\nabla_{\lambda}g_{\mu\nu}\,,
\label{torsion and nonmetricity}
\end{equation}
respectively, where we used the Weitzenb\"{o}ck gauge: Eq.~(\ref{Weitzenbock gauge}). {\it A direct calculation shows $Q^{\rho}{}_{\mu\nu}=0$ by virtue of Eq.~(\ref{relation between spacetime and internal space metric}) in the current convention to the internal-space metric.} Therefore, the torsion only survives in the current theory while automatically vanishing the non-metricity. Whereas, in Symmetric Teleparallel Equivalent to General Relativity (STEGR), we need a special manipulation. In detail, see Sec. II-A in~\cite{Ferraro:2016wht,Tomonari:2024vij}.

Finally, let us introduce the Lagrangian of NGR. In the teleparallel condition and our convention to the internal-space metric, contracting Eq.~(\ref{Generic curvature tensor}) with respect to all possible indices, we first obtain the TEGR Lagrangian density~\cite{Maluf:2013gaa}
\begin{equation}
    \mathcal{L}_{\rm TEGR}=|\theta|\,\mathbb{T}\,=|\theta|\left(-\frac{1}{4}T_{\alpha\mu\nu}T^{\alpha\mu\nu}-\frac{1}{2}T_{\alpha\mu\nu}T^{\mu\alpha\nu}+T^{\alpha}T_{\alpha}\right)\,,
\label{}
\end{equation}
where $|\theta|$ is the determinant of the co-frame field components and the surface term is ignored. Then, the NGR is a quadratic extension of TEGR: the Lagrangian is formulated by a three-parameter theory as follows~\cite{Hayashi:1979qx}:
\begin{equation}
    \mathcal{L}_{NGR}=|\theta|\mathbb{T}=|\theta|\left(c_{1}T_{\alpha\mu\nu}T^{\alpha\mu\nu}+c_{2}T_{\alpha\mu\nu}T^{\mu\alpha\nu}+c_{3}T^{\alpha}T_{\alpha}\right)\,,
\label{}
\end{equation}
where $c_{1}$, $c_{2}$, and $c_{3}$ are are three free parameters. These free parameters bring the abundances of the theory. In fact, as we will see in the next subsection, the NGR is split into nine irreducible classes according to $SO(n)$-symmetry on each leaf in terms of ADM-foliation. 

\subsection{\label{02:02}The NGR Hamiltonian and primary constraints}
From now on, we assume that the dimension of spacetime is four. ADM-foliation~\cite{Deser:1959vvc,Arnowitt:1960es,Arnowitt:1962hi} of a spacetime manifold $(\mathcal{M}\,,g_{\mu\nu}\,,\tilde{\Gamma}^{\rho}{}_{\mu\nu})$ is a diffeomorphism $\sigma\,:\,\mathcal{M}\rightarrow\mathbb{R}\times\mathcal{S}^{3}$ such that it decomposes $\mathcal{M}$ as a disjoint union of hypersurfaces $\Sigma_{t}=\{p\in\mathcal{M}\,|\,\sigma^{*}\tau(p)=t\}$ in a globally hyperbolic manner, which is diffeomorphic to $\{t\}\times\mathcal{S}^{3}$, {\it i.e.,} $\mathcal{M}=\bigsqcup_{t\in\mathcal{T}}$, where $\mathcal{S}^{3}$ denotes a 3-dimensional hypersurface, $\mathcal{T}$ is a time interval of $\mathcal{M}$, $t$ is a time coordinate of $\mathcal{M}$, $\tau$ is a time coordinate of $\mathbb{R}\times\mathcal{S}^{3}$, and $\sigma^{*}$ is the pullback operator of $\sigma$~\cite{Baez:1995sj,Nakahara2003}. Each leaf $\Sigma_{t}$ has a normal vector $n=\xi^{A}e_{A}$, or in the component form, $n^{\mu}=\xi^{A}e^{\mu}{}_{A}$ in a coordinate system of an open set containing the leaf $\Sigma_{t}$, where
\begin{equation}
    \xi^{A}=-\frac{1}{6}\epsilon^{A}{}_{BCD}\theta^{B}{}_{i}\theta^{C}{}_{j}\theta^{D}{}_{k}\epsilon^{ijk}\,,
\label{}
\end{equation}
with satisfying the normalzation condition $\xi^{\mu}\xi_{\mu}=-1$. Namely, each leaf is taken to be a spacelike hypersurface. The normalization condition and the property of Levi-Civita symbols lead to the following algebras:
\begin{equation}
    \eta_{AB}\xi^{A}\xi^{B}=\xi^{A}\xi_{A}=-1,,\quad \eta_{AB}\xi^{A}\theta^{B}{}_{i}=\xi_{A}\theta^{A}{}_{i}=0\,,
\label{Properties of xi}
\end{equation}
respectively. Here, remark that the second algebra above implies that the spatial components of the normal vector $n$ vanish: $n_{i}=\xi_{A}\theta^{A}{}_{i}=0$. This indicates that in this ADM-foliation we can ignore the spatial total divergent term like $D_{i}(\cdots)$ in the spatial integration on a leaf $\Sigma_{t}$ thanks to Stokes' theorem. Then the metric tensor $g_{\mu\nu}$ is decomposed as follows:
\begin{equation}
    g_{\mu\nu}=\begin{bmatrix}
        -\alpha^{2}+\beta^{i}\beta^{j}h_{ij} & \beta_{i} \\
        \beta_{i} & h_{ij}
    \end{bmatrix}\,,\quad
    g^{\mu\nu}=\begin{bmatrix}
    -\frac{1}{\alpha^{2}} & \frac{\beta^{i}}{\alpha^{2}} \\
    \frac{\beta^{i}}{\alpha^{2}} & h^{ij}-\frac{\beta^{i}\beta^{j}}{\alpha^{2}}
    \end{bmatrix}\,,
\label{}
\end{equation}
where $\alpha$ and $\beta^{i}$ are the lapse function and the shift vector, respectively.  $h_{ij}$ is the metric on a leaf $\Sigma_{t}$. According to this metric decomposition, the frame field and the co-frame field are also decomposed into time- and spatial-parts, respectively, as follows: $e_{A}=e_{A}{}^{0}\partial_{0}+e_{A}{}^{i}\partial_{i}$ and $\theta^{A}=\theta^{A}{}_{0}dx^{0}+\theta^{A}{}_{i}dx^{i}=(\alpha\xi^{A}+\beta^{i}\theta^{A}{}_{i})dx^{0}+\theta^{A}{}_{i}dx^{i}$. Therefore, we have
\begin{equation}
    e_{A}{}^{0}=-\frac{1}{\alpha^{2}}\xi_{A}\,,\quad e_{A}{}^{i}=\theta_{A}{}^{i}+\xi_{A}\frac{\beta^{i}}{\alpha}\,,\quad h_{ij}=\eta_{AB}\theta^{A}{}_{i}\theta^{B}{}_{j}\,,
\label{}
\end{equation}
where $\theta_{A}{}^{i} := \eta_{AB}h^{ij}\theta^{B}{}_{j} \neq e_{A}{}^{i}$. We calculate the PB-algebras of the theory based on this variable decomposition in the next section. Remark, here, that the diffeomorphism $\sigma$ can be taken for $\mathcal{M}$ in a {\it global} manner, but once we introduce a coordinate system to express each quantity in terms of the index notation using the spacetime letters, the globality is generically lost. All such quantities are expressed {\it only} in the coordinate system that is taken to be in advance. That is, in the latter case, all the statements are only valid in the {\it local} region covered by the coordinate system. The occurrence of the situation is easy to understand if one considers the existence of coordinate singularities in a generic case.

The configuration space $\mathcal{Q}$ is spanned by the variables: $\alpha$, $\beta^{i}$, $\theta^{A}{}_{i}$, and $\Lambda^{A}{}_{B}$. Canonical momenta with respect to these variables are defined as usual: $\pi_{0}=\delta \mathcal{L}_{NGR}/\delta\dot{\alpha}$, $\pi_{i}=\delta\mathcal{L}_{NGR}/\delta\dot{\beta}^{i}$, $\pi_{A}{}^{i}=\delta\mathcal{L}_{NGR}/\delta\dot{\theta}^{A}{}_{i}$, and $\hat{\pi}^{AB}=\delta\mathcal{L}_{NGR}/\delta a_{AB}$, where $a_{AB}=\eta_{AC}\omega^{C}{}_{B0}=\eta_{C[A}\Lambda^{C}{}_{|D|}(\dot{\Lambda^{-1}})^{D}{}_{B]}$. A direct calculation shows $\pi_{0}=0$ and $\pi_{i}=0$. The momentum $\pi_{A}{}^{i}$ can be decomposed further into $SO(3)$-irreducible components as follows~\cite{Blixt:2018znp}:
\begin{equation}
    \pi_{A}{}^{i}={}^{\mathcal{V}}\pi^{i}\xi_{A}+{}^{\mathcal{A}}\pi^{ji}h_{kj}\theta_{A}{}^{k}+{}^{\mathcal{S}}\pi^{ji}h_{kj}\theta_{A}{}^{k}+{}^{\mathcal{T}}\pi\theta_{A}{}^{i}\,,
\label{}
\end{equation}
where ${}^{\mathcal{V}}\pi^{i}$, ${}^{\mathcal{A}}\pi^{ji}$, ${}^{\mathcal{S}}\pi^{ji}$, and ${}^{\mathcal{T}}\pi$ are the vectorial, anti-symmetric, symmetric trace-free and trace part of the momentum $\pi_{A}{}^{i}$, which are given as follows~\cite{Pati:2022nwi}:
\begin{equation}
\begin{split}
    &{}^{\mathcal{V}}\pi^{i}=-\xi^{A}\pi_{A}{}^{i}\,,\quad{}^{\mathcal{A}}\pi^{ij}=\pi^{[ij]}=-\frac{1}{2}\pi_{A}{}^{i}\theta^{A}{}_{k}h^{jk}+\frac{1}{2}\pi_{A}{}^{j}\theta^{A}{}_{k}h^{ik}\,,\\
    &{}^{\mathcal{S}}\pi^{ij}=\pi^{(ij)}-\frac{1}{3}\pi_{A}{}^{k}\theta^{A}{}_{k}h^{ij}=\frac{1}{2}\pi_{A}{}^{i}\theta^{A}{}_{k}h^{jk}+\frac{1}{2}\pi_{A}{}^{j}\theta^{A}{}_{k}h^{ik}-\frac{1}{3}\pi_{A}{}^{k}\theta^{A}{}_{k}h^{ij}\,,\quad{}^{\mathcal{T}}\pi=\frac{1}{3}\pi_{A}{}^{i}\theta^{A}{}_{i}\,.
\end{split}    
\label{SO3-irreducible canonical momenta}
\end{equation}
The momentum $\hat{\pi}^{AB}$ can be expressed in terms of $\pi_{A}{}^{i}$ and $\theta^{A}{}_{i}$ as follows~\cite{Blixt:2018znp}: $\hat{\pi}^{AB}=-\pi_{C}{}^{i}\eta^{C[B}\theta^{A]}{}_{i}$. As discussed in Sec.~III of Ref.~\cite{Blixt:2018znp}, performing the field redefinition properly, we can show that, as long as we consider the theory in Weitzenb\"{o}ck gauge, the Hamiltonian is independent from $\Lambda^{A}{}_{B}$ and $\hat{\pi}^{AB}$. Therefore, $\mathcal{Q}$ is spanned by the independent variables $\alpha$, $\beta^{i}$, and $\theta^{A}{}_{i}$, and the phase space of $\mathcal{Q}$, $T^{*}\mathcal{Q}$, is spanned by $\alpha$, $\beta^{i}$, $\theta^{A}{}_{i}$, $\pi_{0}$, $\pi_{i}$, and $\pi_{A}{}^{i}$. The total-Hamiltonian of NGR is then derived as follows~\cite{Blixt:2018znp}:
\begin{equation}
\begin{split}
    \mathcal{H} = &\Big({}^\mathcal{V}\mathcal{H} + {}^\mathcal{A}\mathcal{H} + {}^\mathcal{S}\mathcal{H} + {}^\mathcal{T}\mathcal{H} \Big) + D_i[\pi_A{}^i(\alpha \xi^{A}+\beta^{j}\theta^A{}_j )]\\
    &\quad\quad\quad\quad - \alpha \Big(\sqrt{h}\ {}^{3}\mathbb{T} - \xi^A D_i\pi_A{}^i \Big) - \beta^k\Big(T^A{}_{jk}\pi_A{}^j + \theta^A{}_k D_{i}\pi_A{}^i\Big) + {}^{\alpha}\lambda\pi_{0} + {}^{\beta}\lambda^{i}\pi_{i}\,,
\end{split}
\label{}
\end{equation}
where 
\begin{align}
&{}^\mathcal{V} \mathcal{H} =
\begin{cases}
\alpha \sqrt{h}\ \frac{{}^{\mathcal{V}}C_i {}^{\mathcal{V}}C^i}{4 A_{\mathcal{V}}}& \textrm{ for } {}^{\mathcal{V}}A \neq 0\\
\sqrt{h}\ {}^\mathcal{V}\lambda_i {}^{\mathcal{V}}C^i & \textrm{ for } {}^{\mathcal{V}}A = 0\,,
\end{cases}
\quad
{}^\mathcal{A} \mathcal{H} =
\begin{cases}\label{VH and AH}
-\alpha \sqrt{h}\ \frac{{}^{\mathcal{A}}C_{ij} {}^{\mathcal{A}}C^{ij}}{4 A_{\mathcal{A}}}& \textrm{ for } {}^{\mathcal{A}}A \neq 0\\
\sqrt{h}\ {}^\mathcal{A}\lambda_{ij} {}^{\mathcal{A}}C^{ij} & \textrm{ for } {}^{\mathcal{A}}A = 0\,,
\end{cases}\\
&{}^\mathcal{S} \mathcal{H} =
\begin{cases}
-\alpha \sqrt{h}\ \frac{{}^{\mathcal{S}}C_{ij} {}^{\mathcal{S}}C^{ij}}{4 A_{\mathcal{S}}}& \textrm{ for } {}^{\mathcal{S}}A \neq 0\\
\sqrt{h}\ {}^\mathcal{S}\lambda_{ij} {}^{\mathcal{S}}C^{ij} & \textrm{ for } {}^{\mathcal{S}}A = 0\,,
\end{cases}
\quad
{}^\mathcal{T} \mathcal{H} =
\begin{cases}\label{SH and TH}
-\alpha \sqrt{h}\ \frac{3{}^{\mathcal{T}}C {}^{\mathcal{T}}C}{4 A_{\mathcal{T}}}& \textrm{ for } {}^{\mathcal{T}}A \neq 0\\
\sqrt{h}\ {}^\mathcal{T}\lambda {}^{\mathcal{T}}C & \textrm{ for } {}^{\mathcal{T}}A = 0\,,
\end{cases}
\end{align}
where ${}^{\mathcal{A}}\lambda_{ij}$ and ${}^{\mathcal{S}}\lambda_{ij}$ are anti-symmetric and symmetric with respect to the indices $i$ and $j$, respectively, and
\begin{equation}
    {}^{3}\mathbb{T}=c_{1}\eta_{AB}T^A{}_{ij}T^B{}_{kl}h^{ik}h^{jl}+c_{2}\theta_A{}^i \theta_B{}^j T^A{}_{kj}T^B{}_{li}h^{kl}+c_{3}\theta_A{}^i\theta_B{}^j h^{kl}T^A{}_{ki}T^B{}_{lj}\,.
\label{}
\end{equation}
The cases for ${}^{\mathcal{V}}A=0$, ${}^{\mathcal{A}}A=0$, ${}^{\mathcal{S}}A=0$, and/or ${}^{\mathcal{T}}A=0$ are nothing but constraints, and then the variables ${}^\mathcal{V}\mathcal{C}^i$, ${}^\mathcal{A}\mathcal{C}^{ij}$, ${}^\mathcal{S}\mathcal{C}^{ij}$, and ${}^\mathcal{T}\mathcal{C}$ turn to be primary constraints given as follows~\cite{Blixt:2018znp}:
\begin{equation}
\begin{split}
    &{}^\mathcal{V}\mathcal{C}^i := \frac{{}^{\mathcal{V}}\pi^{i}}{\sqrt{h}} - 2  c_{3}T^B{}_{kl} h^{ik}\theta_B{}^l \approx 0\,\quad (A_{\mathcal{V}} = 0)\,,\quad{}^\mathcal{A}\mathcal{C}^{ij} := \frac{{}^\mathcal{A}\pi^{ij}}{\sqrt{h}} - 2  c_{2} h^{li}h^{jk}T^B{}_{kl} \xi_{B} \approx 0\,\quad (A_{\mathcal{A}} = 0)\,,\\
    &{}^\mathcal{S}\mathcal{C}^{ij} :=  \frac{{}^\mathcal{S}\pi^{ij}}{\sqrt{h}} \approx 0\,\quad (A_{\mathcal{S}} = 0)\,,\quad{}^\mathcal{T}\mathcal{C} :=  \frac{{}^\mathcal{T}\pi}{\sqrt{h}} \approx 0\,\quad (A_{\mathcal{T}} = 0)\,.    
\end{split}
\label{SO3-irreducible primary constraints}
\end{equation}
In non-constraint cases, namely for non-vanishing ${}^{\mathcal{V}}A$, ${}^{\mathcal{A}}A$, ${}^{\mathcal{S}}A$, and/or ${}^{\mathcal{T}}A$, the explicit expressions of these variables are given in Ref.~\cite{Blixt:2018znp}. According to the $SO(3)$-irreducible decomposition of the theory, we can classify the theory into nine types as shown in Table~\ref{Types of NGR}~\cite{Blixt:2018znp}. Remark that the constraints $\pi_{0}\approx0$ and $\pi_{i}\approx0$ are common in all types. As shown in the next section, this implies the property of diffeomorphism symmetry of the theory. Remark, finally, that, without imposing the Weitzenb\"{o}ck gauge, a residual symmetry, although which would not be a local one, with respect to the spin connection would arise. It would occur in a way that the configuration space and thus also the phase space are extended into more larger space due to the existence of $\Lambda^{A}{}_{B}$ and $\hat{\pi}^{AB}$. 

\begin{table}[ht!]
    \centering
    \renewcommand{\arraystretch}{1.5}
    \begin{tabular}{ c || c | c }
        Theory & Conditions & $SO(3)$-irreducible primary constraints \\ \hline\hline
        Type 1 & $A_{I}\neq 0 \ \forall I\in \{ \mathcal{V},{\mathcal{A}},{\mathcal{S}},\mathcal{T} \}$ &  No constraint\\ \hline
        Type 2 & $A_{\mathcal{V}}=0$ &  ${}^{\mathcal{V}}\mathcal{C}_{i}\approx0$\\ \hline
        Type 3 & $A_{\mathcal{A}}=0$ &   ${}^{\mathcal{A}}\mathcal{C}_{ij}\approx0$\\ \hline
        Type 4 & $A_{\mathcal{S}}=0$ &   ${}^{\mathcal{S}}\mathcal{C}_{ij}\approx0$\\ \hline
        Type 5 & $A_{\mathcal{T}}=0$ &   ${}^{\mathcal{T}}\mathcal{C}\approx0$\\ \hline
        Type 6 & $A_{\mathcal{V}}=A_{\mathcal{A}}=0$ & ${}^{\mathcal{V}}\mathcal{C}_{i}={}^{\mathcal{A}}\mathcal{C}_{ij}\approx0$\\ \hline
        Type 7 & $A_{\mathcal{A}}=A_{\mathcal{S}}=0$ &  ${}^{\mathcal{A}}\mathcal{C}_{ij}={}^{\mathcal{S}}\mathcal{C}_{ij}\approx0$\\ \hline
        Type 8 & $A_{\mathcal{A}}=A_{\mathcal{T}}=0$ &  ${}^{\mathcal{A}}\mathcal{C}_{ij}={}^{\mathcal{T}}C\approx0$\\ \hline
        Type 9 & $A_{\mathcal{V}}=A_{\mathcal{S}}=A_{\mathcal{T}}=0$ &  ${}^{\mathcal{V}}\mathcal{C}_{i}={}^{\mathcal{S}}\mathcal{C}_{ij}={}^{\mathcal{T}}C\approx0$\\ \hline
    \end{tabular}
    \caption{All types of NGR in the $SO(3)$-irreducible decomposition of canonical momentum. Type 6 is TEGR.}
    \label{Types of NGR}
\end{table}

\section{\label{03}Dirac-Bergmann analysis of Type 2, Type 3, Type 5, and Type 8}
Type 1 and Type 6 are the non-constraint system and TEGR, respectively. The DoF of Type 1 is therefore eight~\cite{Mitric:2019rop}. The DoF of Type 6 is two, and it has been verified by several authors~\cite{Maluf:2000ag,Blagojevic:2000qs,Ferraro:2016wht}. Type 2 was analyzed in Ref.~\cite{Cheng:1988zg}, having six DoF only spelled out in a master thesis mentioned in this article. In the current paper, we focus on the DoF of Type 3, Type 5, and Type 8. In Type 2, we revisit and reconsider it to confirm the previous work.

\subsection{\label{03:01}Common sector: Diffeomorphism symmetry}
The total Hamiltonian of each type is given as follows:
\begin{equation}
\begin{split}
    \mathcal{H}_{\rm Type\,2} =& \sqrt{h}{}^{\mathcal{V}}\lambda_{i}(T^{*}\mathcal{Q}){}^{\mathcal{V}}\mathcal{C}^{i} + D_i[\pi_A{}^i(\alpha \xi^{A}+\beta^{j}\theta^A{}_j )]\\
    & + \alpha \Big( - \sqrt{h}\ {}^{3}\mathbb{T} - \xi^A D_i\pi_A{}^i - \sqrt{h}\ \frac{{}^{\mathcal{A}}\mathcal{C}_{ij} {}^{\mathcal{A}}\mathcal{C}^{ij}}{4 A_{\mathcal{V}}} - \sqrt{h}\ \frac{{}^{\mathcal{S}}\mathcal{C}_{ij} {}^{\mathcal{S}}\mathcal{C}^{ij}}{4 A_{\mathcal{S}}} - \sqrt{h}\ \frac{3{}^{\mathcal{T}}\mathcal{C} {}^{\mathcal{T}}\mathcal{C}}{4 A_{\mathcal{T}}}\Big) + {}^{\alpha}\lambda\pi_{0}\\
    & + \beta^k\Big( - T^A{}_{jk}\pi_A{}^j - \theta^A{}_k D_{i}\pi_A{}^i\Big) + {}^{\beta}\lambda^{i}\pi_{i}\,,
\end{split}
\label{total Hamiltonian of type 2}
\end{equation}

\begin{equation}
\begin{split}
    \mathcal{H}_{\rm Type\,3} =& \sqrt{h}{}^{\mathcal{A}}\lambda_{ij}(T^{*}\mathcal{Q}){}^{\mathcal{A}}\mathcal{C}^{ij} + D_i[\pi_A{}^i(\alpha \xi^{A}+\beta^{j}\theta^A{}_j )]\\
    & + \alpha \Big( - \sqrt{h}\ {}^{3}\mathbb{T} - \xi^A D_i\pi_A{}^i + \sqrt{h}\ \frac{{}^{\mathcal{V}}\mathcal{C}_i {}^{\mathcal{V}}\mathcal{C}^i}{4 A_{\mathcal{V}}} - \sqrt{h}\ \frac{{}^{\mathcal{S}}\mathcal{C}_{ij} {}^{\mathcal{S}}\mathcal{C}^{ij}}{4 A_{\mathcal{S}}} - \sqrt{h}\ \frac{3{}^{\mathcal{T}}\mathcal{C} {}^{\mathcal{T}}\mathcal{C}}{4 A_{\mathcal{T}}}\Big) + {}^{\alpha}\lambda\pi_{0}\\
    & + \beta^k\Big( - T^A{}_{jk}\pi_A{}^j - \theta^A{}_k D_{i}\pi_A{}^i\Big) + {}^{\beta}\lambda^{i}\pi_{i}\,,
\end{split}
\label{total Hamiltonian of type 3}
\end{equation}

\begin{equation}
\begin{split}
    \mathcal{H}_{\rm Type\,5} =& \sqrt{h}{}^{\mathcal{T}}\lambda(T^{*}\mathcal{Q}){}^{\mathcal{T}}\mathcal{C} + D_i[\pi_A{}^i(\alpha \xi^{A}+\beta^{j}\theta^A{}_j )]\\
    & + \alpha \Big( - \sqrt{h}\ {}^{3}\mathbb{T} - \xi^A D_i\pi_A{}^i + \sqrt{h}\ \frac{{}^{\mathcal{V}}\mathcal{C}_i {}^{\mathcal{V}}\mathcal{C}^i}{4 A_{\mathcal{V}}} - \sqrt{h}\ \frac{{}^{\mathcal{S}}\mathcal{C}_{ij} {}^{\mathcal{S}}\mathcal{C}^{ij}}{4 A_{\mathcal{S}}} - \sqrt{h}\ \frac{{}^{\mathcal{A}}\mathcal{C}_{ij} {}^{\mathcal{A}}\mathcal{C}^{ij}}{4 A_{\mathcal{T}}}\Big) + {}^{\alpha}\lambda\pi_{0}\\
    & + \beta^k\Big( - T^A{}_{jk}\pi_A{}^j - \theta^A{}_k D_{i}\pi_A{}^i\Big) + {}^{\beta}\lambda^{i}\pi_{i}\,,
\end{split}
\label{total Hamiltonian of type 5}
\end{equation}

\begin{equation}
\begin{split}
    \mathcal{H}_{\rm Type\,8} =& \sqrt{h}{}^{\mathcal{A}}\lambda_{ij}(T^{*}\mathcal{Q}){}^{\mathcal{A}}\mathcal{C}^{ij} + \sqrt{h}{}^{\mathcal{T}}\lambda{}^{\mathcal{T}}\mathcal{C} + D_i[\pi_A{}^i(\alpha \xi^{A}+\beta^{j}\theta^A{}_j )]\\
    & + \alpha \Big( - \sqrt{h}\ {}^{3}\mathbb{T} - \xi^A D_i\pi_A{}^i + \sqrt{h}\ \frac{{}^{\mathcal{V}}\mathcal{C}_i {}^{\mathcal{V}}\mathcal{C}^i}{4 A_{\mathcal{V}}} - \sqrt{h}\ \frac{{}^{\mathcal{S}}\mathcal{C}_{ij} {}^{\mathcal{S}}\mathcal{C}^{ij}}{4 A_{\mathcal{S}}}\Big) + {}^{\alpha}\lambda\pi_{0}\\
    & + \beta^k\Big( - T^A{}_{jk}\pi_A{}^j - \theta^A{}_k D_{i}\pi_A{}^i\Big) + {}^{\beta}\lambda^{i}\pi_{i}\,,
\end{split}
\label{total Hamiltonian of type 8}
\end{equation}
where $\lambda_{ij}(T^{*}\mathcal{Q})$ denotes the abbreviation of $\lambda_{ij}(\alpha\,,\beta^{i}\,,\theta^{A}{}_{i}\,,\pi_{0}\,,\pi_{i}\,,\pi_{A}{}^{i})$. The total Hamiltonian of another type can be composed of in the same manner, although we do not consider them in the current paper. The fundamental PB-algebra is given as follows:
\begin{equation}
\begin{split}
    &\{\alpha(t\,,\vec{x})\,,\pi_{0}(t\,,\vec{y})\}=\delta^{(3)}(\vec{x}-\vec{y})\,,\quad
    \{\beta^{i}(t\,,\vec{x})\,,\pi_{j}(t\,,\vec{y})\}=\delta^{i}{}_{j}\delta^{(3)}(\vec{x}-\vec{y})\,,\\
    &\{h^{ij}(t\,,\vec{x})\,,\pi_{kl}(t\,,\vec{y})\}=\delta^{i}{}_{(k}\delta^{j}{}_{l)}\delta^{(3)}(\vec{x} - \vec{y})\,,\quad \{\theta^{A}{}_{\mu}(t\,,\vec{x})\,,\pi_{B}{}^{\nu}(t\,,\vec{y})\}=\delta^{A}{}_{B}\delta_{\mu}{}^{\nu}\delta^{(3)}(\vec{x}-\vec{y})\,,
\end{split}
\label{Fundamental PB in NGR}
\end{equation}
where $\pi_{ij}(t\,,\vec{x})$ are the canonical momenta of $h_{ij}$. The PB-algebra of the induced metric $h_{ij}$ and its canonical momentum $\pi_{A}{}^{i}$ is calculated as follows:
\begin{equation}
\begin{split}
    &\{h_{ij}(t\,,\vec{x})\,,\pi_{A}{}^{k}(t\,,\vec{y})\} = 2\eta_{AB}\theta^{B}{}_{(i}\delta_{j)}{}^{k}\delta^{(3)}(\vec{x}-\vec{y})\,.\\
    &\{h^{ij}(t\,,\vec{x})\,,\pi_{A}{}^{k}(t\,,\vec{y})\} = -2h^{im}h^{jl}\eta_{AB}\theta^{B}{}_{(m}\delta_{l)}{}^{k}\delta^{(3)}(\vec{x}-\vec{y})\,.
\end{split}
\label{}
\end{equation}
That is, these variables do not commute in PB-algebra.

The first line in the total Hamiltonian in Eq.~(\ref{total Hamiltonian of type 2}), Eq.~(\ref{total Hamiltonian of type 3}), Eq.~(\ref{total Hamiltonian of type 5}), and Eq.~(\ref{total Hamiltonian of type 8}) governs the constraint structure of the theory since a tedious calculation would show that the second and third lines in each the total Hamiltonian satisfies the following PB-algebra:
\begin{align}
    &\{\phi^{(2)}_{i}(t\,,\vec{x})\,,\phi^{(2)}_{j}(t\,,\vec{y})\}=\left(\phi^{(2)}_{j}(t\,,\vec{x})\partial^{(x)}_{i}-\phi^{(2)}_{i}(t\,,\vec{y})\partial^{(y)}_{j}\right)\delta^{(3)}(\vec{x}-\vec{y})\,,\\
    \quad
    &\{\phi^{(2)}_{i}(t\,,\vec{x})\,,\phi^{(2)}_{0}(t\,,\vec{y})\}=\phi^{(2)}_{0}(t\,,\vec{x})\partial^{(x)}_{i}\delta^{(3)}(\vec{x}-\vec{y})\,,\\
    &\{\phi^{(2)}_{0}(t\,,\vec{x})\,,\phi^{(2)}_{0}(t\,,\vec{y})\}=\left(h^{ij}(t\,,\vec{x})\phi^{(2)}_{j}(t\,,\vec{x})+h^{ij}(t\,,\vec{y})\phi^{(2)}_{j}(t\,,\vec{y})\right)\partial^{(x)}_{i}\delta^{(3)}(\vec{x}-\vec{y})\,,
\label{PB-algebra of diffeomorphism symmetry}
\end{align}
where $\phi^{(2)}_{0}$ and $\phi^{(2)}_{i}$ are the coefficients of the lapse function $\alpha$ and the shift vector $\beta^{i}$ in each the total Hamiltonian, respectively, and these are nothing but the secondary constraint densities of the theory. Then these PB-algebras form the hypersurface deformation algebra~\cite{Dirac:1958sc}. A tedious calculation would also show that $\phi^{(2)}_{0}$ and $\phi^{(2)}_{i}$ commute with all other constraint densities in the theory. Other convenient formulae relating to the fundamental PB-algebras are given in Appendix~\ref{App:01}. This result is consistent with the manifestation of diffeomorphism symmetry of the theory. Therefore, it is the common structure on all types of NGR that the four primary first-class and four secondary first-class constraint densities exist, which are nothing but the generator of diffeomorphism symmetries of the theory. This indicates that the remaining tasks for completing the Dirac-Bergmann analysis of each theory are to investigate the consistency conditions to the first line of the total Hamiltonian of the theory. 

Here, we remark on an important point to make the Dirac-Bergmann analysis valid: the {\it regularity condition} of constraints. Essentially, this condition provides the functional independence of constraints in a way that the constraints form a linearly independent maximal set with coefficients consisting of the phase space variables~\cite{Dirac:1950pj}. However, there is no generic method to derive the regularity condition for a set of constraints. Fortunately, in the case of NGR based on $SO(3)$-irreducible representation of canonical momenta, thanks to its construction, it is clear to specify the doubt of holding the maximal property of the linear independence in ${}^\mathcal{S}\mathcal{C}^{ij} (\propto {}^\mathcal{S}\pi^{ij}) \approx 0\,$. This statement would be justified by the fact that this constraint is trace-free. That is, the trace-free property could be violated in the first-order variation from the constraint surface. In such an irregular case, on one hand, only two out of three diagonal components of ${}^\mathcal{S}\mathcal{C}^{ij}$ are independent due to the necessity of imposing the trace-free property. On the other hand, if this is not the case, {\it i.e.,} in a regular case, the trace-free property holds as strong equality, meaning that all three diagonal components ${}^\mathcal{S}\mathcal{C}^{ij}$ are independent. In the current paper, we restrict to Type 2, Type 3, Type 5, and Type 8 only, and these Types do not contain the constraint ${}^\mathcal{S}\mathcal{C}^{ij} \approx 0\,$. That is, in these Types, the regularity condition always holds, and the Dirac-Bergmann analysis can be applied straightforwardly. In the next paper, other remaining Types, Type 4, Type 7, and Type 9, will be considered, and in these cases, we must take into account an issue due to the dissatisfaction of the regularity condition on the trace-free property in ${}^\mathcal{S}\mathcal{C}^{ij} \propto {}^\mathcal{S}\pi^{ij}\,$. 

Now, we are ready to perform the Dirac-Bergmann analysis~\cite{Dirac:1950pj,Dirac:1958sq,Bergmann:1949zz,BergmannBrunings1949,Bergmann1950,Anderson:1951ta} on each type of NGR. Regarding the importance of the analysis, we perform it firstly from Type 3 and Type 8, then to Type 2 and Type 5. In Type 3, we show that in the consistency conditions we can neglect the spatial total divergent term when solving them with respect to the multipliers. In Type 8, we introduce a new class of constraints that are second-class in terms of density variables but do not determine any Lagrange multipliers. In addition, we find that this new constraint forms a generator of local gauge transformation under the satisfaction of a condition on the Lagrange multiplies. The analysis of Type 2 and Type 5 are performed in a straightforward manner. 

\subsection{\label{03:02}Specific sector in Type 3: Local rotation symmetry and DoF in Type 3}
As mentioned in Sec.~\ref{03:01}, the total Hamiltonian can be split into two parts: a part which governs the specific symmetry depending on each type of NGR and a part which governs the common symmetry of every type of NGR, {\it i.e.,} the diffeomorphism symmetry. Therefore, it is enough for the Dirac-Bergmann analysis on Type 3 of NGR to investigate the following Hamiltonian:
\begin{equation}
    \tilde{\mathcal{H}}_{\rm Type\,3} = \sqrt{h}{}^{\mathcal{A}}\lambda_{ij}{}^{\mathcal{A}}\mathcal{C}^{ij} + D_i[\pi_A{}^i(\alpha \xi^{A}+\beta^{j}\theta^A{}_j )]\,.
\label{specific part of H_type3}
\end{equation}
To proceed with the analysis, we perform the Dirac procedure. To do this, we have to know all PB-algebras among the constraint densities. In the above Hamiltonian, ${}^{\mathcal{A}}\mathcal{C}^{ij}\approx0$ are the primary constraint densities of the theory. The PB-algebras of these constraint densities can be derived from the well-known PB-algebra which is satisfied by the generator of the local Lorentz symmetry given as follows~\cite{Ferraro:2016wht}:
\begin{equation}
    \{\mathcal{C}_{(TEGR)}^{IJ}(t\,,\vec{x})\,,\mathcal{C}_{(TEGR)}^{KL}(t\,,\vec{y})\} = \left({\eta^{JL}}\,\mathcal{C}_{(TEGR)}^{IK}+{\eta^{IK}}\,\mathcal{C}_{(TEGR)}^{JL}-{\eta^{JK}}\,\mathcal{C}_{(TEGR)}^{IL}-{\eta^{IL}}\,\mathcal{C}_{(TEGR)}^{JK}\right)\delta^{(3)}(\vec{x}-\vec{y})\,,
\label{local Lorentz symmetry}
\end{equation}
where $\mathcal{C}_{(TEGR)}^{IJ}\approx0$ are the primary constraint density of TEGR. Restricting the algebra to a leaf $\Sigma_{t}$, we obtain
\begin{equation}
    \{{}^\mathcal{A}\mathcal{C}^{ij}(t\,,\vec{x})\,,{}^\mathcal{A}\mathcal{C}^{kl}(t\,,\vec{y})\} = \frac{2}{\sqrt{h}}\left(\delta^{j[l}{}^{\mathcal{A}}\mathcal{C}^{k]i} + \delta^{i[k}{}^{\mathcal{A}}\mathcal{C}^{l]j}\right)\delta^{(3)}(\vec{x}-\vec{y})\,.
\label{PB-algebra of AC and AC}
\end{equation}
Therefore, ${}^{\mathcal{A}}\mathcal{C}^{ij}\approx0$ are classified as first-class. This algebra implies that if the PB-algebra of ${}^{\mathcal{A}}\mathcal{C}^{ij}\approx0$ with the total divergent term in Eq.~(\ref{specific part of H_type3}) commutes on the ground of imposing the constraint densities, then the Dirac-procedure stops, and the DoF of Type 3 of NGR is determined. 

Here, on one hand, notice that the consistency conditions of the constraint densities should be integrated over on each leaf $\Sigma_{t}$ to be mathematically well-defined equations since all consistency conditions are density equations~\cite{DAmbrosio:2023asf}. On the other hand, as mentioned in Sec.~\ref{02:02}, a spatial total divergent term: $D_{i}(\cdots)$ vanishes when integrating over on a leaf $\Sigma_{t}$ in the specific ADM-foliation given in Sec.~\ref{02:02}. In addition, if the theory satisfies the property of diffeomorphism symmetry, then without loss of generality, a result which is derived by applying the vanishing property in the specific ADM-foliation holds in any ADM-foliation~\cite{Tomonari:2023wcs}. The problem we are facing now is just in this case: The spatial total divergent term $\{{}^{\mathcal{A}}\mathcal{C}^{ij}(t\,,\vec{x})\,,D^{(y)}_k[\pi_A{}^k(\alpha \xi^{A}+\beta^{k}\theta^A{}_k )](t\,,\vec{y})\} = D^{(y)}_k\{{}^{\mathcal{A}}\mathcal{C}^{ij}(t\,,\vec{x})\,,\pi_A{}^k(\alpha \xi^{A}+\beta^{k}\theta^A{}_k )(t\,,\vec{y})\}$ vanishes in the present ADM-foliation. The explicit calculation of the consistency condition ${}^\mathcal{A}\mathcal{C}^{ij}(t\,,\vec{x})\approx0$ is performed as follows:
\begin{equation}
\begin{split}
        &\frac{d}{dt}\int_{\Sigma_{t}}{}^\mathcal{A}\mathcal{C}^{ij}(t\,,\vec{x})dx^{3} = \left\{\int_{\Sigma_{t}}{}^\mathcal{A}\mathcal{C}^{ij}(t\,,\vec{x})dx^{3}\,,\int_{\Sigma_{t}}\tilde{H}_{\rm Type\,3}(t\,,\vec{y})dy^{3}\right\}\\
        &\approx\int_{\Sigma_{t}}dx^{3}\int_{\Sigma_{t}}dy^{3}\,\sqrt{h}\,\left[{}^\mathcal{A}\lambda_{kl}\frac{2}{\sqrt{h}}\left(\delta^{j[l}{}^{\mathcal{A}}\mathcal{C}^{k]i} + \delta^{i[k}{}^{\mathcal{A}}\mathcal{C}^{l]j}\right)\delta^{(3)}(\vec{x}-\vec{y}) + \left\{{}^{\mathcal{A}}\mathcal{C}^{ij}(t\,,\vec{x})\,,D^{(y)}_k[\pi_A{}^k(\alpha \xi^{A}+\beta^{k}\theta^A{}_k )](t\,,\vec{y})\right\}\right]\\
        &= \int_{\Sigma_{t}}dx^{3}\left[2\,{}^\mathcal{A}\lambda_{kl}\left(\delta^{j[l}{}^{\mathcal{A}}\mathcal{C}^{k]i} + \delta^{i[k}{}^{\mathcal{A}}\mathcal{C}^{l]j}\right)\right] + \int_{\Sigma_{t}}dx^{3}\int_{\Sigma_{t}}dy^{3}\,\sqrt{h}\,D^{(y)}_k\left\{{}^{\mathcal{A}}\mathcal{C}^{ij}(t\,,\vec{x})\,,[\pi_A{}^k(\alpha \xi^{A}+\beta^{k}\theta^A{}_k )](t\,,\vec{y})\right\}\\
        &= \int_{\Sigma_{t}}dx^{3}\left[2\,{}^\mathcal{A}\lambda_{kl}\left(\delta^{j[l}{}^{\mathcal{A}}\mathcal{C}^{k]i} + \delta^{i[k}{}^{\mathcal{A}}\mathcal{C}^{l]j}\right)\right] + \int_{\Sigma_{t}}dx^{3}\,\sqrt{h}\,n_{k}(t\,,\vec{y})\left\{{}^{\mathcal{A}}\mathcal{C}^{ij}(t\,,\vec{x})\,,[\pi_A{}^k(\alpha \xi^{A}+\beta^{k}\theta^A{}_k )](t\,,\vec{y})\right\}\\
        &= \int_{\Sigma_{t}}dx^{3}\left[2\,{}^\mathcal{A}\lambda_{kl}\left(\delta^{j[l}{}^{\mathcal{A}}\mathcal{C}^{k]i} + \delta^{i[k}{}^{\mathcal{A}}\mathcal{C}^{l]j}\right)\right]\approx0\,,
\end{split}
\label{}
\end{equation}
where we denoted ``$D^{(x)}_{i}$'' by the covariant derivative with respect to the spatial coordinate ``$\vec{x}$'' on a leaf $\Sigma_{t}$. In the second and last line we used the constraint density: ${}^{\mathcal{A}}\mathcal{C}^{ij}\approx0$. In the fourth and fifth lines we used Stokes' theorem and the property of the normal vector, $n_{i}=\xi_{A}\theta^{A}{}_{i}=0$, given in Sec.~\ref{02:02}. In addition, as discussed in Sec.~\ref{03:01}, Type 3 of NGR is a diffeomorphism invariant theory. Therefore, without loss of generality, we can neglect the spatial total divergent term in the total Hamiltonian Eq.~(\ref{total Hamiltonian of type 3}) and count out the DoF of Type 3 of NGR. That is, the DoF of Type 3 of NGR is $(16\times2 - 8\times2 - 3\times2)/2=5$. 

Type 3 of NGR has first-class constraint densities only. This means that the constraint densities form a generator of local gauge symmetry. The generator can be composed of the linear combination of the first-class constraint densities with the coefficients of functions~\cite{Sugano:1986xb,Sugano:1989rq,Sugano:1991ke,Sugano:1991kd,Sugano:1991ir}. For the diffeomorphism symmetry, the generator is composed as follows:
\begin{equation}
    \mathcal{G}_{\rm Diffeo} = g_{(1)}^{\mu}\phi^{(1)}_{\mu}+g_{(2)}^{\mu}\phi^{(2)}_{\mu}\,,
\label{Gauge genrator for Diffeomorphism}
\end{equation}
where $\phi^{(1)}_{\mu}:=(\pi_{0}\,,\pi_{i})$, $\phi^{(2)}_{\mu}=(\phi^{(2)}_{0}\,,\phi^{(2)}_{i})$ and, $g^{(1)}_{\mu}$ and $g^{(2)}_{\mu}$ are a set of arbitrary functions. For the three dimensional rotation symmetry, {\it i.e.,} the symmetry on each leaf $\Sigma_{t}$, the generator is composed by
\begin{equation}
    \mathcal{G}_{\rm Type\,3} = g^{i}L_{i}\,,\quad L_{i}=\frac{1}{2}\epsilon_{ijk}\,{}^{\mathcal{A}}\mathcal{C}^{jk}\,,
\label{Gauge generator for Type3}
\end{equation}
where $g^{i}$ are a set of arbitrary functions. In fact, the variation of the total-Hamiltonian Eq.~(\ref{total Hamiltonian of type 3}) becomes as follows:
\begin{equation}
    \delta_{\rm Diffeo}\tilde{\mathcal{H}}_{\rm Type\,3} = \left\{\tilde{\mathcal{H}}_{\rm Type\,3}\,,\mathcal{G}_{\rm Diffeo}\right\} = -\mathcal{L}_{\mathcal{G}_{\rm Diffeo}}\tilde{\mathcal{H}}_{\rm Type\,3}\approx0\,
\label{}
\end{equation}
and
\begin{equation}
    \delta_{\rm Type\,3}\tilde{\mathcal{H}}_{\rm Type\,3} = \left\{\tilde{\mathcal{H}}_{\rm Type\,3}\,,\mathcal{G}_{\rm Type\,3}\right\} = -\mathcal{L}_{\mathcal{G}_{\rm Type\,3}}\tilde{\mathcal{H}}_{\rm Type\,3}\approx0\,
\label{}
\end{equation}
as desired, where ``$\mathcal{L}_{X}$'' denotes the Lie derivative with respect to the vector field ``$X$''. These properties indicate the consistency of the theory: Type 3 of NGR is invariant under the diffeomorphism in the spacetime and rotation transformation in the internal-space. 

\subsection{\label{03:03}Specific sector in Type 8: New behavior of constraints and DoF in Type 8}
The part of governing the proper symmetry in Type 8 of NGR is given by
\begin{equation}
    \tilde{\mathcal{H}}_{\rm Type\,8} = \sqrt{h}{}^{\mathcal{A}}\lambda_{ij}(T^{*}\mathcal{Q}){}^{\mathcal{A}}\mathcal{C}^{ij} + \sqrt{h}{}^{\mathcal{T}}\lambda(T^{*}\mathcal{Q}){}^{\mathcal{T}}{C} + D_i[\pi_A{}^i(\alpha \xi^{A} + \beta^{j}\theta^A{}_j)]\,.
\label{specific part of H_type8}
\end{equation}
In this type of NGR, the primary constraint densities are ${}^{\mathcal{A}}\mathcal{C}^{ij}\approx0$ and ${}^{\mathcal{T}}\mathcal{C}\approx0$. As mentioned in Sec.~\ref{03:02}, we can neglect the total divergent term, {\it i.e.,} the last term in Eq.~(\ref{specific part of H_type8}) in this analysis by virtue of the diffeomorphism symmetry. Therefore, the PB-algebras among ${}^{\mathcal{A}}\mathcal{C}^{ij}$ and ${}^{\mathcal{T}}\mathcal{C}$ determined the DoF of the theory.

The PB-algebra of ${}^{\mathcal{T}}\mathcal{C}$ itself is commute with one another as follows:
\begin{equation}
    \{{}^{\mathcal{T}}\mathcal{C}(t\,,\vec{x})\,,{}^{\mathcal{T}}\mathcal{C}(t\,,\vec{y})\}=0\,.
\label{PB-algebra of TC and TC}
\end{equation}
The PB-algebra of ${}^{\mathcal{A}}\mathcal{C}^{ij}$ itself is already computed in Sec.~\ref{03:02}: Eq.~(\ref{PB-algebra of AC and AC}). The PB-algebra of ${}^{\mathcal{T}}\mathcal{C}$ and ${}^{\mathcal{A}}\mathcal{C}^{ij}$ are calculated as follows:
\begin{equation}
\begin{split}
    \{{}^{\mathcal{T}}\mathcal{C}(t\,,\vec{x})\,,{}^{\mathcal{A}}\mathcal{C}^{ij}(t\,,\vec{y})\} = 
    & \frac{1}{3}\frac{1}{\sqrt{h(t\,,\vec{x})}}{}^{\mathcal{A}}\mathcal{C}^{ij}(t\,,\vec{x})\delta^{(3)}(\vec{x}-\vec{y})\\
    &\quad\quad\quad\quad + \frac{2}{3}\frac{1}{\sqrt{h(t\,,\vec{x})}}\left(-2c_{2}h^{il}(t\,,\vec{y})h^{jk}(t\,,\vec{y})\xi_{A}(t\,,\vec{y})\theta^{A}{}_{[k}(t\,,\vec{x})\partial^{(y)}_{l]}\delta^{(3)}(\vec{x}-\vec{y})\right)\,,    
\end{split}
\label{PB-algebra of TC and AC}
\end{equation}
where we denoted ``$\partial^{(x)}_{i}$'' as the partial derivative with respect to the spatial coordinate ``$\vec{x}$'' on a leaf $\Sigma_{t}$. This PB-algebra can be derived by using the formulae given in Appendix~\ref{App:01}. Remark that in the second term \\$\xi_{A}(t\,,\vec{y})\theta^{A}{}_{[k}(t\,,\vec{x})\partial^{(y)}_{l]}\delta^{(3)}(\vec{x}-\vec{y})$ does not vanish since $\xi_{A}(t\,,\vec{y})$ and $\theta^{A}{}_{i}(t\,,\vec{x})$ are given at two different points. That is, the second property given in Eq.~(\ref{Properties of xi}) can be applied to $\xi_{A}$ and $\theta^{A}{}_{i}$ only at the same point: $\xi_{A}(t\,,\vec{x})\theta^{A}{}_{i}(t\,,\vec{x})=0$. This difference leads to an unfamiliar property of constraints in that of the point particle systems: a property only in field theories arises. Before investigating the classification of the constraints and the determination of the multipliers, let us consider this property. Assume that a vector $F^{A}$ in the internal-space, which is orthogonal to $\xi^{A}$, exists. Then, integrating $\xi_{A}(t\,,\vec{x})F^{A}(t\,,\vec{y})\partial^{(y)}_{i}\delta^{(3)}(\vec{x}-\vec{y})$ over with respect to both the spatial coordinates $\vec{x}$ and $\vec{y}$ on a leaf, we can perform the following calculation:
\begin{equation}
\begin{split}
    &\int_{\Sigma_{t}}dx^{3}\int_{\Sigma_{t}}dy^{3}\,\sqrt{h}\,\xi_{A}(y)F^{A}(x)\partial^{(y)}_{i}\delta^{(3)}(\vec{x} - \vec{y})\\
    &=\int_{\Sigma_{t}}dx^{3}\int_{\Sigma_{t}}dy^{3}\left[\partial^{(y)}_{i}\left(\,\sqrt{h}\,\xi_{A}(y)F^{A}(x)\delta^{(3)}(\vec{x} - \vec{y})\right) - \partial^{(y)}_{i}\left(\,\sqrt{h}\,n_{A}(y)F^{A}(x)\right)\delta^{(3)}(\vec{x} - \vec{y})\right]\\
    &\underset{y \in \partial\Sigma_{t}}{=} n_{i}(y)\int_{\Sigma_{t}}dx^{3}\,\sqrt{h}\,\xi_{A}(y)F^{A}(x)\delta^{(3)}(\vec{x} - \vec{y}) - \int_{\Sigma_{t}}dx^{3}\partial^{(x)}_{i}\left(\,\sqrt{h}\,\xi_{A}(x)F^{A}(x)\right)\\
    &\underset{x\,,y \in \partial\Sigma_{t}}{=}\,\sqrt{h}\,\left[n_{i}(y)\xi_{A}(y)F^{A}(y) - n_{i}(x)\xi_{A}(x)F^{A}(x)\right]\\
    &\underset{x\,,y \in \partial\Sigma_{t}}{=}\,\sqrt{h}\,\left[0 - 0\right]=0\,,
\end{split}
\label{integration of xi^A theta_A}
\end{equation}
where $y\in\partial\Sigma_{t}$ and $x\,,y\in\partial\Sigma_{t}$ denote evaluating the line on the boundary of $\Sigma_{t}$: $\partial\Sigma_{t}$. Notice that we do not need the property of vanishing $n_{i}$ in the above calculation. That is, {\it in smeared variables, the orthogonal property of $\xi^{A}$ and $F^{A}$ at the same point implies that the term $\xi_{A}(t\,,\vec{x})F^{A}(t\,,\vec{y})\partial^{(y)}_{i}\delta^{(3)}(\vec{x}-\vec{y})$ does not affect the classification either a constraint is first-class or second-class.} Therefore, ${}^{\mathcal{A}}\mathcal{C}^{ij}\approx0$ and ${}^{\mathcal{T}}\mathcal{C}\approx0$ are, on one hand, classified as second-class in terms of density variables due to the non-vanishing second term in the PB-algebra of Eq.~(\ref{PB-algebra of TC and AC}) under the imposition of the constraint densities. On the other hand, in terms of smeared variables,\footnote{
A smeared variable of $F^{\cdots}{}_{\cdots}$ with respect to the constraint ${}^{\mathcal{X}}\mathcal{C}^{\cdots}{}_{\cdots}$ is defined as follows:
\begin{equation}
    {}^{\mathcal{X}}\,C[F^{\cdots}{}_{\cdots}]=\int_{\Sigma_{t}}dx^{3}\,F^{\cdots}{}_{\cdots}\,{}^{\mathcal{X}}\mathcal{C}^{\cdots}{}_{\cdots}\,,
\label{}
\end{equation}
where ``$\cdots$'' in ${}^{\cdots}{}_{\cdots}$ of $F$ and of ${}^{\mathcal{X}}\mathcal{C}$ and ${}^{\mathcal{X}}C$ are dummy indices. $\mathcal{X}$ is taken to be $\mathcal{T}$ or $\mathcal{A}$.  
}
these constraints are classified as first-class:
\begin{equation}
    \left\{{}^{\mathcal{T}}C[\,{}^{\mathcal{T}}F]\,,{}^{\mathcal{A}}C^{ij}[\,{}^{\mathcal{A}}F]\right\} = {}^{\mathcal{A}}C^{ij}[\,\frac{1}{3\sqrt{h}}{}^{\mathcal{T}}F\,{}^{\mathcal{A}}F\,] + {}^{\mathcal{A}}C^{ij}[\,{}^{\mathcal{T}}F\mathcal{L}_{{}^{\mathcal{T}}\mathcal{C}}{}^{\mathcal{A}}F\,] -{}^{\mathcal{T}}C[\,{}^{\mathcal{A}}F\mathcal{L}_{{}^{\mathcal{A}}\mathcal{C}^{ij}}{}^{\mathcal{T}}F\,] \,,
\label{smeared PB-algebra of TC and AC}
\end{equation}
where ${}^{\mathcal{T}}F$ and ${}^{\mathcal{A}}F$ are arbitrary test functions. In particular, the test function ${}^{\mathcal{A}}F$ can be replaced by a vector ${}^{\mathcal{A}}F^{i}$ or a tensor ${}^{\mathcal{A}}F^{ij}$ with contracting each index with that of ${}^{\mathcal{A}}C_{ij}$. The modification of Eq.~(\ref{smeared PB-algebra of TC and AC}) is trivial. One can confirm this statement by replacing the dummy vector ``$F^{A}$'' in the internal-space and the index ``$i$'' by ``$h^{il}h^{jk}\theta^{A}{}_{k}$'' and ``$l$'', respectively, and anti-symmetrizing with respect to ``$k$'' and ``$l$'', since the existence of the spatial indices on a leaf does not affect the calculation in Eq.~(\ref{integration of xi^A theta_A}). As mentioned in Sec.~\ref{03:02}, since the consistency conditions are density equations, it should be integrated over with respect to all spatial coordinate variables on a common leaf $\Sigma_{t}$ to be well-defined mathematically. Then we can show that the consistency conditions of ${}^{\mathcal{A}}\mathcal{C}^{ij}\approx0$ and ${}^{\mathcal{T}}\mathcal{C}\approx0$ do not determine any multipliers. Therefore, the analysis stops here, and the DoF is $(16\times2 - 8\times2 - 4)/2 = 6$ in terms of density variables. In terms of smeared variables, the DoF is $(16\times2 - 8\times2 -4\times2)/2 = 4$. Notice, here, that {\it 
the constraints do not determine any Lagrange multipliers whether or not the actual classification of the constraints is first-class.} This situation leads us to the next problem: Which the DoF should be the correct one? The clue to resolve this situation is the gauge invariance of the theory. 

As mentioned in Sec.~\ref{03:02}, the gauge generator can be composed of the linear combination of the first-class constraint densities with arbitrary functions as its coefficients. Since the diffeomorphism symmetry holds in the same manner as Type 3 of NGR, we focus on the specific symmetry of Type 8 here. Let us consider the following gauge generators:
\begin{equation}
    \mathcal{G}_{\rm Type\,8} = {}^{\mathcal{T}}g\,{}^{\mathcal{T}}\mathcal{C} + {}^{\mathcal{A}}g_{ij}\,{}^{\mathcal{A}}\mathcal{C}^{ij}\,
\label{Gauge generator for Type8 in density variables}
\end{equation}
in terms of density variables, where ${}^{\mathcal{T}}g$ and ${}^{\mathcal{A}}g_{ij} = - {}^{\mathcal{A}}g_{ji}$ are arbitrary functions, and 
\begin{equation}
    G_{\rm Type\,8} = \int_{\Sigma_{t}}dx^{3}\,\mathcal{G}_{\rm Type\,8} = {}^{\mathcal{T}}C[\,{}^{\mathcal{T}}g\,] + {}^{\mathcal{A}}C[\,{}^{\mathcal{A}}g\,]\,
\label{Gauge generator for Type8 in smeared variables}
\end{equation}
in terms of smeared variables, where in the latter equation we abbreviated the indices of ${}^{\mathcal{A}}g_{ij}$ and just expressed as ``${}^{\mathcal{A}}g$'' since the indices are contracted over with corresponding those of ${}^{\mathcal{A}}C^{ij}$. Remark that Eq.~(\ref{Gauge generator for Type8 in density variables}) is the generator of a {\it global} symmetry even if ${}^{\mathcal{T}}\mathcal{C}\approx0$ and ${}^{\mathcal{A}}\mathcal{C}^{ij}\approx0$ are second-class~\cite{Sugano:1989rq}. Let us investigate first the easier one: In the latter case, we can easily verify 
\begin{equation}
    \delta_{\rm smeared\,Type\,8}\tilde{\mathcal{H}}_{\rm Type\,8} = \{\tilde{\mathcal{H}}_{\rm Type\,8}\,,G_{\rm Type\,8}\} = -\mathcal{L}_{G_{\rm Type\,8}}\tilde{\mathcal{H}}_{Type\,8} \approx0\,.
\label{Variation of H8 in smeared variables}
\end{equation}
Namely, the theory is gauge invariant. In the former one, we obtain the following relation:
\begin{equation}
    \delta_{\rm Type\,8}\tilde{\mathcal{H}}_{\rm Type\,8} \approx \frac{2}{3}\frac{1}{\sqrt{h(t\,,\vec{x})}}\left(-2c_{2}h^{il}(t\,,\vec{y})h^{jk}(t\,,\vec{y})\xi_{A}(t\,,\vec{y})\theta^{A}{}_{[k}(t\,,\vec{x})\partial^{(y)}_{l]}\delta^{(3)}(\vec{x}-\vec{y})\right)\left(\,{}^{\mathcal{T}}g\,{}^{\mathcal{A}}\lambda_{ij} - \,{}^{\mathcal{A}}g_{ij}\,{}^{\mathcal{T}}\lambda\,\right)\,.
\label{Variation of H8 in density variables}
\end{equation}
If we impose the following condition to the multipliers and the coefficients of the gauge generator then the theory turns to be gauge invariant:
\begin{equation}
    \,{}^{\mathcal{T}}g\,{}^{\mathcal{A}}\lambda_{ij} - \,{}^{\mathcal{A}}g_{ij}\,{}^{\mathcal{T}}\lambda\, = 0\,.
\label{Condition of gauge invariance of H8 in density variables}
\end{equation}
That is, if this condition holds then the theory is {\it local} gauge invariant under the gauge transformation generated by Eq.~(\ref{Gauge generator for Type8 in density variables}) although ${}^{\mathcal{T}}\mathcal{C}\approx0$ and ${}^{\mathcal{A}}\mathcal{C}^{ij}\approx0$ are classified as second-class in terms of density variables. As mentioned in Ref.~\cite{Sugano:1989rq}, second-class constraints also generate a {\it global} gauge symmetry. In contrast to this statement, {\it in Type 8 of NGR, the satisfaction of Condition~(\ref{Condition of gauge invariance of H8 in density variables}) leads to a {\it local} gauge symmetry even in the case that the generator is composed of second-class constraint densities.} That is, this result implies that the Dirac-Bergmann analysis in the use of density variables and of smeared variables does not always coincide. Remark that Condition~(\ref{Condition of gauge invariance of H8 in density variables}) just restrict the solution space of the gauge functions, ${}^{\mathcal{T}}g$ and ${}^{\mathcal{A}}g_{ij}$, and any multipliers are determined unless these functions are identified according to the method provided in Ref.~\cite{Sugano:1989rq}. A crucial issue would be to provide a proof of the existence of {\it non-}empty solution space, but we leave it for our future investigation, which will be discussed from a more generic perspective. 

Armed with the consideration in the last paragraph, we should derive the DoF is Case 1: four if Condition~(\ref{Condition of gauge invariance of H8 in density variables}) holds; Case 2: six if Condition~(\ref{Condition of gauge invariance of H8 in density variables}) does not hold. Based on the consideration so far, let us generalize the property of the constraints discussed above and define a new class of constraints by the following statement: If a second-class constraint density does not determine any Lagrange multiplier(s) and composes a basis of a gauge generator under the satisfaction of a set of conditions to the Lagrange multiplier(s) and the coefficient(s) of the gauge generator, then let the constraint density be a ``{\it semi-}''{\it first-class constraint density}. 
This peculiar characteristics is ascribed to the nature of field theories, which is absent in point particle systems. P. M. A. Dirac established a generic theory of canonical constraint analysis in point particle systems only, thus he could not encounter this nature in field theories.~\footnote{
Dirac applied his analysis to GR to consider quantum gravity based on canonical quantization in Ref.~\cite{Dirac:1958sc}. However, in GR, the complicated situation that occurs in Type 8 of NGR did not appear.}
The fundamental property of semi-first-class constraint densities will be investigated in a sequel paper. 

\subsection{\label{03:04}Specific sector in Type 2: DoF in Type 2}
The part of the total-Hamiltonian in Type 2 we should consider is given as follows:
\begin{equation}
    \tilde{\mathcal{H}}_{\rm Type\,2} = \sqrt{h}{}^{\mathcal{V}}\lambda_{i}(T^{*}\mathcal{Q}){}^{\mathcal{V}}\mathcal{C}^{i} + D_i[\pi_A{}^i(\alpha \xi^{A}+\beta^{j}\theta^A{}_j )]\,.
\label{specific part of H_type2}
\end{equation}
The primary constraint density of the theory is ${}^{\mathcal{A}}\mathcal{C}^{ij}\approx0$. The PB-algebra is derived from Eq.~(\ref{local Lorentz symmetry}) as follows: 
\begin{equation}
    \{{}^{\mathcal{V}}\mathcal{C}^{i}(t\,,\vec{x})\,,{}^{\mathcal{V}}\mathcal{C}^{j}(t\,,\vec{y})\} = {}^{\mathcal{A}}\mathcal{C}^{ij}\delta^{(3)}(\vec{x}-\vec{y})\,.
\label{PB-algebra of VC and VC}
\end{equation}
In Type 2, since ${}^{\mathcal{A}}C^{ij}$ are not constraint density, ${}^{\mathcal{V}}\mathcal{C}^{i}\approx0$ are classified as second-class. The consistency condition of the primary constraint density becomes as follows: 
\begin{equation}
    \begin{bmatrix}
        {}^{\mathcal{V}}\dot{C}^{1} \\
        {}^{\mathcal{V}}\dot{C}^{2} \\
        {}^{\mathcal{V}}\dot{C}^{3}
    \end{bmatrix}
    =\int_{\Sigma_{t}}dx^{3}\,\sqrt{h}\,\begin{bmatrix}
        0 & {}^{\mathcal{A}}\mathcal{C}^{12} & {}^{\mathcal{A}}\mathcal{C}^{13} \\
        {}^{\mathcal{A}}\mathcal{C}^{21} & 0 & {}^{\mathcal{A}}\mathcal{C}^{23} \\
        {}^{\mathcal{A}}\mathcal{C}^{31} & {}^{\mathcal{A}}\mathcal{C}^{32} & 0
    \end{bmatrix}
    \begin{bmatrix}
        {}^{\mathcal{A}}\lambda_{1} \\
        {}^{\mathcal{A}}\lambda_{2} \\
        {}^{\mathcal{A}}\lambda_{3}
    \end{bmatrix}
    :\approx0\,.
\label{consistency conditions of primary constraints in type 2}
\end{equation}
where ``$:\approx$'' denotes the imposition as a condition in weak equality and the integration acts on each row of the equations. In particular, Eq.~(\ref{consistency conditions of primary constraints in type 2}) can be solved when those in terms of density variables are solved in this case. Since ${}^{\mathcal{A}}C^{ij} \approx 0$ are anti-symmetric with respect to the indices ``$i$'' and ``$j$'', ${\rm det}\,({}^{\mathcal{A}}C^{ij}) = 0$ and its rank is two. Therefore, only two out of the three multipliers are determined. Without loss of generality, the multipliers ${}^{\mathcal{A}}\lambda_{1}$ and ${}^{\mathcal{A}}\lambda_{2}$ can be expressed in terms of ${}^{\mathcal{A}}\lambda_{3}$ by
\begin{equation}
    {}^{\mathcal{A}}\lambda_{1} = \frac{{}^{\mathcal{A}}\mathcal{C}^{23}}{{}^{\mathcal{A}}\mathcal{C}^{12}}{}^{\mathcal{A}}\lambda_{3}\,\quad {}^{\mathcal{A}}\lambda_{2} = -\frac{{}^{\mathcal{A}}\mathcal{C}^{13}}{{}^{\mathcal{A}}\mathcal{C}^{12}}{}^{\mathcal{A}}\lambda_{3}\,
\label{}
\end{equation}
and these expressions satisfy ${}^{\mathcal{A}}\mathcal{C}^{31}{}^{\mathcal{A}}\lambda_{1} + {}^{\mathcal{A}}\mathcal{C}^{32}{}^{\mathcal{A}}\lambda_{2} = 0$ automatically. Then the specific part of the total-Hamiltonian, Eq.~(\ref{specific part of H_type2}), is written as follows:
\begin{equation}
    \tilde{\mathcal{H}}_{\rm Type\,2} = {}^{\mathcal{A}}\lambda_{3}\,\frac{2\sqrt{h}}{{}^{\mathcal{A}}\mathcal{C}^{12}}{}^{\mathcal{V}}\mathcal{C}^{i}L_{i}\,,\quad L_{i}=\frac{1}{2}\epsilon_{ijk}\,{}^{\mathcal{A}}\mathcal{C}^{jk}\,.
\label{modified specific part of H_type2}
\end{equation}
Then we can show that $\dot{C}^{i}\approx0$ for all $i=1\,,2\,,3$ by straightforward computations. Therefore, we have three primary second-class constraint densities. However, the total number has to be even~\cite{Shanmugadhasan:1973ad,Maskawa:1976hw,Dominici:1979bg,Tomonari:2023vgg}. This means that we have at least one more second-class constrain density: It would be 
\begin{equation}
    {}^{\mathcal{V}}\mathcal{C} := L_{i}\,{}^{\mathcal{V}}\mathcal{C}^{i}\approx0\,.
\label{}
\end{equation}
This constraint density satisfies its consistency condition. The PB-algebras among ${}^{\mathcal{V}}\mathcal{C}$ and ${}^{\mathcal{V}}\mathcal{C}^{i}$ are calculated as follows:
\begin{equation}
    \{{}^{\mathcal{V}}\mathcal{C}\,,{}^{\mathcal{V}}\mathcal{C}\}\approx0\,\quad \{{}^{\mathcal{V}}\mathcal{C}^{i}\,,{}^{\mathcal{V}}\mathcal{C}\}\approx{}^{\mathcal{A}}\mathcal{C}^{ij}L_{j}\,.
\label{}
\end{equation}
Therefore, ${}^{\mathcal{V}}\mathcal{C}\approx0$ is classified as second-class. This result is consistent with the statement given in Ref.~\cite{Cheng:1988zg}. The Dirac-procedure stops here, and the DoF is $(16\times2 - 8\times2 - 4)/2 = 6$. In Ref.~\cite{Cheng:1988zg}, the authors indicated that the consistency conditions may bifurcate into a tree structure. In this point, we briefly discuss in Sec.~\ref{04}.

\subsection{\label{03:05}Specific sector in Type 5: DoF in Type 5}
The part of the total-Hamiltonian in Type 5 we should consider is given as follows:
\begin{equation}
    \tilde{\mathcal{H}}_{\rm Type\,5} = \sqrt{h}{}^{\mathcal{T}}\lambda(T^{*}\mathcal{Q}){}^{\mathcal{T}}\mathcal{C} + D_i[\pi_A{}^i(\alpha \xi^{A}+\beta^{j}\theta^A{}_j )]\,.
\label{specific part of H_type5}
\end{equation}
The primary constraint density of the theory is ${}^{\mathcal{T}}\mathcal{C}\approx0$\,. The PB-algebra of the primary constraint density is already given in Eq.~(\ref{PB-algebra of TC and TC}). Therefore, the primary constraint density satisfies the consistency condition automatically and is classified as first-class. The Dirac-procedure stops here. The DoF is $(16\times2 - 8\times2 - 1\times2)/2=7$. 

\subsection{\label{03:05} Interpretation of DoFs}
Let us interpret the DoF in each Type. The configuration space consists of the lapse function $\alpha$, the shift vector $\beta^{i}$, and the co-frame field components $\theta^{A}{}_{i}$. The first two quantities are decomposed into two scalar DoFs and two vector DoFs since the Helmholtz decomposition theorem suggests that the shift vector can be decomposed into rotation-free and divergence-free vector parts that can be decomposed further into scalar and vector fields, respectively~\cite{Hu:2022anq}. The co-frame field components can be translated into the induced metric in the spacetime foliation, $h_{ij}$, via the relation of $h_{ij} = \theta^{A}{}_{i}\theta^{B}{}_{j}\eta_{AB}$ with the local Lorentz Symmetry (LS) in a one-to-one manner. The induced metric is decomposed further into two tensor DoFs, two vector DoFs, and two scalar DoFs. Hence, the theory can have up to two tensor DoFs, four scalar DoFs, four vector DoFs. The total number of these DoFs is ten and this matches the total number of the spacetime metric tensor components. In many cases including GR (Type 6 in NGR) and NGR, however, since the sector composed of the lapse function and the shift vector are fully constraint in primary constraints and reduces the total DoFs by two scalar and two vector DoFs, the total DoFs in the metric sector become two tensor DoFs, two vector DoFs, and two scalar DoFs: eight DoFs in total. 

If the diffeomorphism symmetry holds such as GR (, or equivalently TEGR up to boundary terms in variation principle,) and NGR then the hypersurface deformation algebra reduces two scalar DoFs and two vector DoFs from the total DoFs, the remaining two tensor DoFs that correspond to the massless spin-2 field in Minkowski background. However, if the diffeomorphism symmetry is broken, then some of the two scalar and two vector DoFs can be revived. In generic massive gravity~\cite{Comelli:2012vz}, for instance, the mass term breaks this symmetry, and five DoFs (two tensor, one scalar, and two vector) and one ghost (one scalar) DoF so-called Boulware-Deser ghost arise, which correspond to massive spin-2 field in Minkowski background.~\footnote{
In dRGT massive gravity, thanks to a symmetry protecting it from the Boulware-Deser ghost, the theory turns out to be ghost-free.
} However, in NGR, since the diffeomorphism symmetry holds, the metric sector provides two tensor DoFs only. Thus, in NGR, the debatable point is the possibility of breaking the local LS only~\cite{Mariz:2022oib}. That is, if this symmetry is broken then the one-to-one property in $h_{ij} = \theta^{A}{}_{i}\theta^{B}{}_{j}\eta_{AB}$ is lost, and then the extra DoFs appear up to six. Remark that the violation of the local LS is also possible to lead to massive gravity~\cite{Rubakov:2004eb,Dubovsky:2004sg,Rubakov:2008nh}. Generically, just the canonical analysis cannot determine what modes propagate in the theory. In Ref.~\cite{Bahamonde:2024zkb}, propagating modes on Minkowski background are investigated, providing that possible composition of those modes for each Type. In the final section~\ref{04}, we summarized our results together with possible propagating modes as in Table~\ref{Table:ghostsI and fullDoF}. However, the relation of these propagating modes to massive gravity cannot be unveiled only from the canonical analysis, and we leave it for future investigations.

From the view point of the canonical analysis, the upper bound of DoFs generated from the symmetry breaking only is predictable in each Type of NGR. Combining with the results in Ref.~\cite{Bahamonde:2024zkb}, we may anticipate possible composition of propagating modes for each Type in a higher-order perturbation theory or/and the linear perturbation theory on another background. Let us estimate what propagating mode(s) arises, focusing only on Type 1, Type 2, and Type 3, as follows. (The cases of other Types can be considered in the same manner.)
In Type 1, since the local LS is completely lost, the theory obtains six extra DoFs adding to two tensor DoFs. Thus, the total DoF is eight. In terms of Propagating Modes (PMs), these eight DoFs should be composed of two tensor, one pseudo-scalar, one scalar, two pseudo-vector, and two vector modes. 
In Type 2, since the generators ${}^{\mathcal{A}}C_{ij}$ are absent, thus the three extra DoFs arise. In terms of PMs, one pseudo-scalar and two pseudo-vector modes are possible to arise. In addition, the generators ${}^{\mathcal{V}}C_{i}$ are possible to revive at least one extra DoF since second-class constraints are not {\it static} but {\it stationary}. In terms of PMs, one scalar mode is possible to arise. Thus, the extra DoFs in Type 2 should be up to four, and the total DoFs are up to six. In terms of PMs, one scalar, one pseudo-scalar, and two pseudo-vector modes are possible to arise, adding to the two tensor modes. 
In Type 3, the extra DoFs corresponding to the absence of ${}^{\mathcal{V}}C_{i}$ appear: three, since $SO(3)$-symmetry, which is a subgroup of the local LS, holds. Thus, the total DoFs are up to five. In terms of PMs, one scalar and two vector modes are possible to arise, adding to the two tensor modes. These total DoFs are equal to or larger than the results of the canonical analysis in each Type, respectively, thus all the ingredients are consistent. 

\section{\label{04}Conclusions}
In this work we have performed the Hamiltonian analysis of a subset of NGR theories, which are of greater physical interest. Depending on (non)presence of primary constraints, NGR can be classified into nine distinct theories, where Type 1 is the generic theory and Type 6 corresponds to TEGR, both being well-studied and, hence, excluded from the analysis of this work. Furthermore, we restrict ourselves to theories with a propagating spin-2 field, which exclude Type 4, Type 7, and Type 9 \cite{Bahamonde:2024zkb}. The remaining physically interesting theories are Type 2, Type 3, Type 5, and Type 8, where the last two might not be consistent with cosmological observations since the torsion scalar vanishes for simple FLRW backgrounds. Furthermore, in the case of Type 5 and Type 8 the conformal mode of the spin-2 field is not predictable \cite{Golovnev:2023ddv}.

At the beginning of this work, in Sec.~\ref{02}, TEGR and NGR were revisited from the viewpoint of gauge approach to gravity, and then NGR in the $SO(3)$-irreducible representation of canonical momenta was reviewed. After that, in Sec.~\ref{03}, applying the Dirac-Bergmann analysis, we investigated the full DoF of NGR, focusing on Type 2, Type 3, Type 5, and Type 8. We unveiled that the full DoF of these types are, in this order, six, five, seven, and either four under the specific condition of Lagrange multipliers or six in a generic case, respectively. That is, in Type 8 a bifurcation occurred, while in Type 2, we did not confirm a bifurcation which is indicated in Ref.~\cite{Cheng:1988zg}. In particular, in Sec.~\ref{03:03}, we found in Type 8 a novel new behavior of second-class constraint densities, in which the constraints compose the generator of the gauge transformation of the theory not only in the smeared variables but also in the density variables on the ground of the satisfaction of a set of proper conditions of the Lagrange multipliers. It is well-known knowledge that a set of first-class constraint densities forms a generator of gauge transformation, but our work showed that so does for that of second-class constraint densities under the satisfaction of proper conditions on Lagrange multipliers. In this context, we should not overlook the property of Lagrange multipliers known as `{\it Lambda Symmetry}' proposed and discussed in Refs.~\cite{Blagojevic:2000pi,Blagojevic:2023fys}, which reduces the gauge symmetry of the theory. In our work, we showed that if Eq.~(\ref{Condition of gauge invariance of H8 in density variables}) is satisfied, then Type 8 turns into a {\it locally} gauge invariant theory, and the gauge functions are restricted by the multipliers and vice versa. 

We should mention the bifurcation in consistency conditions further. This sort of bifurcation was addressed in the first attempt at the Dirac-Bergmann analysis of NGR presented in Ref.~\cite{Cheng:1988zg}, which is caused by the violation of (local) Lorentz symmetry. In this work, the analysis of Type 2 was performed, and it was clarified that the theory has four second-class constraints only, which coincides with our result, and there might be a bifurcation, although this is not confirmed in our analysis. It may be ascribed to the difference in the manipulation of consistency conditions. That is, in our analysis, we required the stationary condition among all constraints only, but in Ref.~\cite{Cheng:1988zg} the condition is alternated by a static condition; all constraints should vanish not only in the first-order time derivatives (stationary condition) but also in arbitrarily higher-order derivatives (static condition). The latter conditions are stronger than ours which were originally introduced by P. M. A. Dirac~\cite{Dirac:1950pj,Dirac:1958sq}.

We should remark that a correct manipulation of the bifurcation is not so simple. One can confirm this point from a lesson of the case of $f(T)$-gravity. The authors in Ref.~\cite{Blagojevic:2020dyq} found five independent sectors which contain the result presented in Ref.~\cite{Li:2011rn}, that is, five DoF in the most generic sector. In Ref.~\cite{Ferraro:2018tpu}, the authors found a new sector that is not provided in Ref.~\cite{Blagojevic:2020dyq}. Thus, there was a controversy in the analysis~\cite{Blixt:2020ekl}, and in Sec.~5.1 of Ref.~\cite{Tomonari:2023wcs} this controversy was reconciled that the result in Ref.~\cite{Ferraro:2018tpu} is the generic sector of (s4) in Ref.~\cite{Blagojevic:2020dyq} by investigating the relation of the consistency conditions in these two analyses. Remark here that this bifurcation is the different one mentioned in Ref.~\cite{DAmbrosio:2023asf,Tomonari:2023wcs}, which is caused by the violation of diffeomorphism symmetry.

Recently, the linear perturbation of NGR around Minkowski background spacetime was investigated~\cite{Bahamonde:2024zkb}, and the work revealed that the propagating DoF of each type of NGR in that momentum representation. Table~\ref{Table:ghostsI and fullDoF} below is the summary of the recent work combining with our work. We find that Type 1, Type 5, and Type 6 are healthy branches from the strong coupling perspective. Since Type 1 and Type 5 suffers from the existence of the ghost modes, we cannot use this branch to describe physical phenomena. Type 6 is nothing but TEGR, and this branch is equivalent to GR up to surface terms in the Lagrangian density. 
\begin{table}[ht!]
    \centering
    \renewcommand{\arraystretch}{1.5}
    \begin{tabular}{ c || c | c | c | c }
        Theory &  Parameter space & Linear DoF & Ghost-free condition & Nonlinear DoF \\ 
        \hline
        \hline
        Type 1 & Generic & 8: $(h^{TT}_{ij},\chi,\psi,U_i,V_i)$ & Impossible & 8 \\ \hline
        Type 2 & $c_{\rm vec}=-c_{\rm ten}$ & 3: $(h^{TT}_{ij},\chi)$ & $c_{\rm ten} > \frac{4}{9}c_{\rm axi},\quad\text{and}\quad c_{\rm ten}>0$ & 6 \\ \hline
        Type 3 & $c_{\rm ten}=\frac{4}{9}c_{\rm axi}$ & 3: $(h^{TT}_{ij},\psi)$ & $(c_{\rm vec}>0,\, c_{\rm ten}>0)$ or $( 0< c_{\rm ten}< -c_{\rm vec} )$ & 5\\ \hline
        Type 5 & $c_{\rm vec}=0$ & 7: $(h^{TT}_{ij},\chi,U_i,V_i)$ &  Impossible & 7\\ \hline
        Type 6 & $c_{\rm ten}=\frac{4}{9}c_{\rm axi}\,,\,\, c_{\rm vec} = - c_{\rm ten}$ & 2: $h^{TT}_{ij}$ &  $c_{\rm ten}>0$ & 2 \\ \hline
        Type 8 & $c_{\rm ten}=\frac{4}{9}c_{\rm axi}\,,\,\, c_{\rm vec} = 0$ & 2: $h^{TT}_{ij}$ & $c_{\rm ten}>0$ & 6 (Generic) or 4 (Special)\\ \hline
    \end{tabular}
    \caption{Propagating linear and nonliear DoF in different types of NGR and necessary conditions for the theory being ghost-free. ``Special'' denotes the case that occurs only under the satisfaction of a set of specific conditions on Lagrange multipliers, whereas ``Generic'' denotes the case without any conditions. Type 6 is TEGR. For the propagating DoFs in detail, see Ref.~\cite{Bahamonde:2024zkb}, referring to linear perturbations around a Minkowski background, for which the result also have been verified using the xAct package PSALTER \cite{Barker:2024juc}.}
    \label{Table:ghostsI and fullDoF}
\end{table} 

For future perspectives, we list the following five directions; 1, It is important to investigate the more detailed property of the novel behavior of second-class constraint densities, which was found in Type 8 in the current paper, and the precise feature of the condition Eq.~(\ref{Condition of gauge invariance of H8 in density variables}). These topics would invite us not only to new perspectives on the Lambda symmetry but also to a deeper understanding of the Dirac-Bergmann analysis.; 2, A possibility of the bifurcation in Type 2 addressed in Ref.~\cite{Cheng:1988zg} should be scrutinized in more detail. In the application to cosmology, we need to unveil the propagating DoFs together with the ghost-free regime in the flat and non-flat FLRW spacetime, not only in the Minkowskian case, and clarify the existence of the strong couplings. In addition, investigations of strong couplings in the UV-regime should also be undertaken as a mandatory work since the propagating DoF(s) might be alternated in this regime.; 3, To coincide the predictions of NGR with the observations today in the solar scale, it is necessary to invent a mechanism such as the Vainstein mechanism known in the scalar-tensor theories of gravity~\cite{Vainshtein:1972sx,Babichev:2013usa} for manipulating properly the extra DoFs that are absent in GR.; 4, Taking the considerations in Sec.~\ref{03:05} into account, the identification of each propagating DoF in Table~\ref{Table:ghostsI and fullDoF} in terms of the DoFs generated by violating the local Lorentz symmetry should be investigated in more detail~\cite{Mariz:2022oib}. In particular, a possibility of massive gravity caused by the violation should be addressed~\cite{Rubakov:2004eb,Dubovsky:2004sg,Rubakov:2008nh}. The perturbation theories of NGR in terms of (co-)frame field would also provide a deeper understanding of this issue~\cite{Bahamonde:2022ohm,Izumi:2012qj}. It also is valuable to extend perturbations into higher-orders or different backgrounds.; 5, It would be interesting to revisit the work of \cite{Blixt:2023qbg}, by extending it to include constructions built on Type 3 NGR, which was previously thought to be plagued by ghost instabilities. To unveil the nature of all five DoFs in this Type, particularly, the consideration of $f(T)$-gravity given in Ref.~\cite{Golovnev:2020nln} would be applicable and provides a great insight. All of these works are open for future work. 

\begin{acknowledgments}
KT and DB would like to thank Alejandro Jiménez Cano, Keisuke Izumi, and Sebastian Bahamonde for insightful and fruitful discussions. KT would like to thank the cosmology theory group in Institute of Science Tokyo for supporting my work, in particular, professor Teruaki Suyama.
\end{acknowledgments}

\appendix
\section{\label{App:01}Complemental PB-algebras}
The PB-algebras relating to the volume element of the induced metric are given by
\begin{equation}
\begin{split}
    &\{\sqrt{h(t\,,\vec{x})}\,,\pi_{A}{}^{i}(t\,,\vec{y})\}=\sqrt{h(t\,,\vec{x})}e_{A}{}^{i}(t\,,\vec{x})\delta^{(3)}(\vec{x}-\vec{y})\,,\\
    &\{\frac{1}{\sqrt{h(t\,,\vec{x})}}\,,\pi_{A}{}^{i}(t\,,\vec{y})\}=-\frac{1}{\sqrt{h(t\,,\vec{x})}}e_{A}{}^{i}(t\,,\vec{x})\delta^{(3)}(\vec{x}-\vec{y})\,.
\end{split}
\label{}
\end{equation}
The PB-algebras relating to the calculation in Type 8 of NGR are given by
\begin{equation}
\begin{split}
    &\{{}^{\mathcal{T}}\pi(t\,,\vec{x})\,,{}^{\mathcal{A}}\pi^{ij}(t\,,\vec{y})\}=-\frac{2}{3}{}^{\mathcal{A}}\pi^{ij}(t\,,\vec{x})\delta^{(3)}(\vec{x}-\vec{y})\,,\\
    &\{{}^{\mathcal{T}}\pi(t\,,\vec{x})\,,h^{ij}(t\,,\vec{y})\}=\frac{2}{3}h^{ij}(t\,,\vec{x})\delta^{(3)}(\vec{x}-\vec{y})\,,\\
    &\{{}^{\mathcal{T}}\pi(t\,,\vec{x})\,,h_{ij}(t\,,\vec{y})\}=-\frac{2}{3}h_{ij}(t\,,\vec{x})\delta^{(3)}(\vec{x}-\vec{y})\,,\\
    &\{{}^{\mathcal{T}}\pi(t\,,\vec{x})\,,T^{A}{}_{ij}(t\,,\vec{y})\}=\frac{2}{3}\theta^{A}{}_{[i}(t\,,\vec{x})\partial^{(y)}_{j]}\delta^{(3)}(\vec{x}-\vec{y})\,,\\
    &\{{}^{\mathcal{T}}\pi(t\,,\vec{x})\,,\xi_{A}(t\,,\vec{y})\}=-\xi_{A}(t\,,\vec{x})\delta^{(3)}(\vec{x}-\vec{y})\,,\\
    &\{\frac{1}{\sqrt{h(t\,,\vec{x})}}\,,{}^{\mathcal{A}}\pi^{ij}(t\,,\vec{y})\}=0\,,\\
    &\{\frac{1}{\sqrt{h(t\,,\vec{x})}}\,,{}^{\mathcal{T}}\pi(t\,,\vec{y})\}=-\frac{1}{\sqrt{h(t\,,\vec{x})}}\delta^{(3)}(\vec{x}-\vec{y})\,.
\end{split}
\label{PB-algebras in Type 8}
\end{equation}

\bibliographystyle{utphys}
\bibliography{Bibliography}
\end{document}